\documentclass[11pt,a4paper]{article}

\usepackage{jcappub}
\usepackage{epstopdf}
\DeclareGraphicsExtensions{.pdf,.png,.jpg}
\usepackage{amsmath}
\usepackage{amsfonts}
\usepackage{amssymb}
\usepackage{url}

\title{Scaling cosmology with variable dark-energy equation of state }

\author[a]{ David R. Castro, }
\author[a,b]{ Hermano~Velten}
\author[a]{and Winfried Zimdahl}

\affiliation[a]{Universidade Federal do Esp\'{\i}rito Santo,
Departamento
de F\'{\i}sica\\
Av. Fernando Ferrari, 514, Campus de Goiabeiras, CEP 29075-910,
Vit\'oria, Esp\'{\i}rito Santo, Brazil}

\affiliation[b]{Fakult\"at f\"ur Physik, Universit\"at Bielefeld, Postfach 100131, 33501 Bielefeld, Germany}

\emailAdd{drodriguez-ufes@hotmail.com}
\emailAdd{velten@physik.uni-bielefeld.de}
\emailAdd{winfried.zimdahl@pq.cnpq.br}

\abstract{Interactions between dark matter and dark energy which result in a power-law behavior (with respect to the cosmic scale factor) of the ratio between the energy densities of the dark components (thus generalizing the $\Lambda$CDM model) have been considered as an attempt to alleviate the cosmic coincidence problem phenomenologically. We generalize this approach by allowing for a variable equation of state for the dark energy within the CPL-parametrization. Based on analytic solutions for the Hubble rate and using the Constitution and Union2 SNIa sets, we present a statistical analysis and classify different interacting and non-interacting models according to the Akaike ($AIC$) and the Bayesian ($BIC$) information criteria. We do not find noticeable evidence for an alleviation of the coincidence problem with the mentioned type of interaction.
}


\begin{document}
\maketitle


\section{Introduction}

Despite of the many efforts of the past decade, the physical nature of the cosmological dark sector remains mysterious.
The preferred model for the current cosmological dynamics is still the $\Lambda$CDM scenario, although there is an ongoing discussion (see, e.g., \cite{nieuwenh}) on  shortcomings of this model which, however, became a reference model also for competing approaches. Much work has been done to study deviations from the $\Lambda$CDM model. Usually, this implies introducing new parameters which are then constrained by the ever increasing number of available observational data.
In this way one tries to clarify, e.g., whether the effective dark-energy equation-of-state (EoS) is really described by a constant parameter and whether this parameter is really equal to -1 as required by a true cosmological constant.
Most investigations impose limits on potential deviations from $\Lambda$CDM, but generally the latter turns out to be  consistent with the data and even remains the preferred option. For recent results see, e.g., \cite{ratra} and \cite{montesano}.

The $\Lambda$CDM model is characterized by a pressureless dark matter component with an energy density $\rho_m$ which decays with the third power $a^{3}$ of the cosmic scale factor $a$ and a constant energy density $\rho_{\Lambda}$, associated with the cosmological constant. Consequently, for the ratio of the energy densities one has
$\frac{\rho_m}{\rho_{\Lambda}} \propto a^{-3}$. Given the huge range of values for $a$ during the cosmic evolution, a currently measured ratio $\frac{\rho_m}{\rho_{\Lambda}}= {\cal
O}(1)$ seems to single out the present epoch as a very special period. This is known as the cosmic coincidence problem. There have been several attempts which tried to uncover a dynamical mechanism to make this coincidence in a sense natural.
One of the possibilities that have been studied over the last years is to admit an interaction between dark matter and dark energy and to check its influence on the ratio of the energy densities of both components.
In this paper we consider an approach that relies on a phenomenological relation for the ratio of the energy densities of dark matter, $\rho_{m}$, and a dynamical dark energy component with density $\rho_{x}$,
\begin{eqnarray}
\frac{\rho_m}{\rho_x}=ra^{-\xi}\ ,
\label{ratio}
\end{eqnarray}
where the scale factor  $a$ of the Robertson-Walker metric was normalized to a present value $a(t_0) =a_0 = 1$, and $r$ is the energy-density ratio at the present time. The power $\xi$ is a constant, non-negative parameter. A behavior of the type (\ref{ratio}) was suggested by Dalal {\it et al.} \cite{dalal} in order to address the coincidence problem on a phenomenological basis.
For an equation of state $p_x = -\rho_x $ of the dark-energy component, a value $\xi =3$ amounts to the
$\Lambda$CDM model.  A value $\xi =0$ represents a stationary ratio
$\frac{\rho_m}{\rho_x} = {\rm const}$.
According to \cite{dalal}, the
deviation of the parameter $\xi$ from $\xi =0$ quantifies the
severity of the coincidence problem. Any solution with a scaling
parameter $\xi < 3$ will make the coincidence problem less severe. Investigations along this line include
\cite{scaling,somasri,alcaniz}. Here we generalize previous studies by admitting a time varying equation of state for the dark-energy component. Using the widely applied CPL parametrization \cite{cpl} for the equation of state parameter, we obtain analytic solutions for the special cases $\xi =1$ and $\xi =3$. In a subsequent statistical analysis our results are confronted with the SNIa data from the Constitution \cite{Hicken} and Union2 \cite{Amanullah} samples and complementary information from the baryon acoustic oscillations (BAO) scale \cite{Eisenstein} and the position of the first acoustic peak in the CMB TT spectrum \cite{komatsu}. Similarities and differences to alternative studies in the literature are pointed out. The results are assessed according
to the Akaike ($AIC$) and Bayesian ($BIC$) information criteria.
We find that none of the models considered here can really compete with the $\Lambda$CDM model. However, this conclusion is partially based on a prior which is adapted to the  $\Lambda$CDM model so that the latter seems naturally preferred. Therefore, the mentioned result may not yet be the final answer. But at least at the moment, there is no evidence that the coincidence problem can be alleviated by an approach  based on (\ref{ratio}) with a corresponding interaction.

The paper is organized as follows. In section \ref{interaction} we present our interacting dark-energy model and obtain analytic solutions for the Hubble rates for two special cases. A corresponding statistical analysis is the subject of section \ref{analysis}. A summary and a discussion of our results are given in section \ref{discussion}.

\section{A suitable interaction}
\label{interaction}

\subsection{General relations}

Let us assume  the cosmic medium to be dynamically dominated by cold dark matter (subindex $m$) and dark energy (subindex $x$). Additionally, we take into account a baryon component (subindex $b$). Then
\begin{equation}
\rho = \rho_{m} + \rho_{x} + \rho_{b}\
\qquad  {\rm and}\qquad
p = p_{m} + p_{x} + p_{b}\
\label{2}
\end{equation}
are the total energy density and the total pressure, respectively.
The components are assumed to possess the equations of state
\begin{equation}
p _{m} \ll \rho _{m}\ ,\quad p _{b} \ll \rho _{b} \quad {\rm and} \quad
p _{x} = w\rho _{x}\ .
\label{3}
\end{equation}
An interaction between both dark components may be described by the set
\begin{equation}
\dot{\rho}_{m} + 3H \rho_{m} = Q
\label{balm}
\end{equation}
and
\begin{equation}
\dot{\rho}_{x}
+ 3H \left(1+\omega\right)\rho _{x} = -Q \ .
\label{balx}
\end{equation}
For the separately conserved baryon component we have
\begin{equation}
\dot{\rho}_{b} + 3H \rho_{b} = 0 \quad\Rightarrow \quad \rho_{b} \propto a^{-3}\ .
\label{balb}
\end{equation}
A specific expression for the interaction term can be derived by considering the
time evolution of the ratio $\frac{\rho _{m}}{\rho _{x}}$,
\begin{equation}
\left(\frac{\rho _{m}}{\rho _{x}} \right)^{\displaystyle \cdot}
= \frac{\rho _{m}}{\rho _{x}}
\left[\frac{\dot{\rho }_{m}}{\rho _{m}}
- \frac{\dot{\rho }_{x}}{\rho _{x}}\right] = 3H \frac{\rho _{m}}{\rho _{x}}
\left[\omega
+ \frac{\rho _{m} + \rho _{x}}{\rho _{m}\rho _{x}}\frac{Q}{3H}\right]\ .
\label{dotr2}
\end{equation}
Inserting here the ansatz (\ref{ratio}) and solving for the interaction term
yields
\begin{equation}
\frac{Q}{3H \rho_m}=- \frac{\omega+\frac{\xi}{3}}{r a^{-\xi}+1} \ .
\label{Q}
\end{equation}
Eq.~(\ref{Q}) demonstrates that by choosing a suitable
interaction between both components, we may produce any desired
scaling behavior of the energy densities \cite{scaling}.
The uncoupled case,
corresponding to $Q = 0$, is given by $\frac{\xi }{3} + \omega = 0$.
The $\Lambda$CDM model corresponds to $\omega = -1$ and $\xi = 3$. Generally, interacting models are parameterized by deviations from $\xi =-3\omega$. Any solution which deviates from $\xi = -3w$ represents a testable,
non-standard cosmological model.
For $\xi > 0$ the expression (\ref{Q}) becomes very small for $a\ll 1$. Consequently,
the interaction is not relevant at high redshifts. This guarantees the existence of an early matter-dominated epoch.
An energy transfer from dark energy to dark matter, i.e. $Q>0$, requires $\frac{\xi }{3} + \omega < 0$.

All relations so far are valid for EoS parameters that are not necessarily constant. In order to take into account a variable EoS parameter, we resort to the CPL parametrization \cite{cpl}
\begin{equation}
 \omega=\omega_0+\omega_1 (1-a)\ ,
 \label{cpl}
\end{equation}
where $\omega_0$ and $\omega_1$ are constant parameters.
Then the condition for $Q>0$ depends on $a$. We have $\omega_{0} + \omega_{1} + \frac{\xi }{3} < 0$ for $a\ll 1$,
$\omega_{0}+ \frac{\xi }{3} < 0$ for $a = 1$ and $\omega_{0} - \omega_{1}a + \frac{\xi }{3} < 0$ for $a\gg 1$. But for the latter limit this kind of parametrization does not seem to be useful.

The dimensionless interaction quantity (\ref{Q}) can be written
\begin{equation}
\frac{Q}{3H \rho_m}=-\left(\frac{a^{\xi}}{r+a^{\xi}}\right) (\omega_0+\omega_1+\frac{\xi}{3}) + \omega_1\left( \frac{a^{\xi+1}}{r+a^{\xi}}\right)\ .
\label{Qwow1}
\end{equation}
In the following we consider separately the analytically solvable cases $\xi=1$ and $\xi=3$. The latter is expected to test primarily deviations from the $\Lambda$CDM model due to a time-varying equation of state for the dark energy, the former
should provide information about the feasibility to alleviate the coincidence problem with the help of a suitable interaction.

To situate our analysis properly, we start with the non-interacting case in the following subsection.

\subsection{The non-interacting case}

If the dark components are uncoupled, equivalent to $Q=0$ in the balances (\ref{balm}) and (\ref{balx}), the resulting square of the Hubble rate, after implementing the CPL parametrization, is \cite{bueno}
\begin{equation}\label{Hbueno}
\left[\frac{H(a)}{H_0}\right] ^2 = \frac{\left(\Omega_{m0} + \Omega_{b0}\right)}{a^3}
+ \frac{\left[1 - \left(\Omega_{m0} + \Omega_{b0}\right)\right]}{a^{3\left(1+\omega_{0} + \omega_{1}\right)}}\exp\left[3\omega_1 \left(a-1\right)\right]\ ,
\end{equation}
where
\begin{equation}
\Omega_{b0} = \frac{8\pi G \rho_{b0}}{3 H_{0}^{2}}\ , \quad \Omega_{m0} = \frac{8\pi G \rho_{m0}}{3 H_{0}^{2}} \quad \mathrm{and} \quad \Omega_{x0} = \frac{8\pi G \rho_{x0}}{3 H_{0}^{2}}
 \label{omegas}
\end{equation}
with $\Omega_{b0}+\Omega_{m0}+\Omega_{x0}=1$.

\subsection{The case $\xi=1$}

Under this condition, integration of the balance equations (\ref{balm}), (\ref{balx}) and (\ref{balb})  provides us with
\begin{equation}
\rho_m=\rho_{m0}a^{-3}\left[  \frac{r+a}{r+1}\right] ^{-3(\frac{1}{3}+\omega_0+\omega_1+\omega_1 r)}\exp\left[ 3\omega_1(a-1)\right] \ ,\qquad \rho_x=\frac{\rho_m}{r}\,a
\label{rm}
\end{equation}
and
\begin{equation}
 \rho_b=\rho_{b0} a^{-3}\ ,
 \label{rb}
\end{equation}
respectively. Taking into account
\begin{equation}
r= \frac{\Omega_{m0}}{\Omega_{x0}} = \frac{\Omega_{m0}}{1-\Omega_{b0}-\Omega_{m0}}\ ,
\end{equation}
we find
\begin{equation}
 \left[ \frac{H(a)}{H_0}\right] ^2=\frac{\Omega_{b0}}{a^3}+ \frac{\left( 1-\Omega_{b0}\right)^{\left(1+3y \right)}}{\left(1-\Omega_{b0}+\Omega_{m0}z\right)^{3y}a^{3\left(1+y  \right)}}
\exp\left[3\omega_1 \left(a-1\right)\right]\ ,
\label{H1}
\end{equation}
for the square of the Hubble rate,
where
\begin{equation}\label{y}
y \equiv \omega_0+\omega_1\left(\frac{1-\Omega_{b0}}{1-\Omega_{b0}-\Omega_{m0}}\right)\ .
\end{equation}
Disregarding the baryon component and setting $\omega_1 = 0$ we recover the results of \cite{scaling}.

\subsection{The case $\xi=3$}
For this case the analytic solution for the matter density is
\begin{eqnarray}
\rho_{m}=\rho_{m0}a^{-3}\left[\frac{r+a^3}{r+1}\right]^{-(\omega_0+\omega_1+1)}
\left[\frac{r^{1/3}+1}{r^{1/3}+a}
\sqrt{\frac{r^{2/3}-r^{1/3}a+a^2}{r^{2/3}-r^{1/3}+1}}\right]^{\omega_1r^{1/3}}
e^{\left[3\omega_1(a-1) + \Delta\right]}\ ,\nonumber\\
\end{eqnarray}
where
\begin{equation}
\Delta=3\omega_1\frac{r^{1/3}}{\sqrt{3}}
\left(\arctan\left[\frac{r^{1/3}-2a}{\sqrt{3}r^{1/3}}\right]-\arctan\left[\frac{r^{1/3}-2}{\sqrt{3}r^{1/3}}\right]\right)\ .
\end{equation}
With the abbreviation $u \equiv \omega_0+\omega_1+1$, the corresponding Hubble rate is given by
\begin{eqnarray}
\left[ \frac{H(a)}{H_0}\right] ^2&=&\frac{\Omega_{b0}}{a^{3}}+\Omega_{m0}\frac{r+a^{3}}{r a^{3}}
\left[\frac{r+1}{r+a^{3}}\right]^{u}\left[\frac{r^{1/3}+1}{r^{1/3}+a}
\sqrt{\frac{r^{2/3}-r^{1/3}a+a^2}{r^{2/3}-r^{1/3}+1}}\right]^{\omega_1r^{1/3}}e^{\left[3\omega_1(a-1) + \Delta\right]}\ .\nonumber\\
\label{H3}
\end{eqnarray}
In the following we shall investigate the cosmological dynamics based on the Hubble rates (\ref{Hbueno}),  (\ref{H1}) and (\ref{H3}).

\begin{figure}[!h]
\centering
\includegraphics[width=0.5\textwidth]{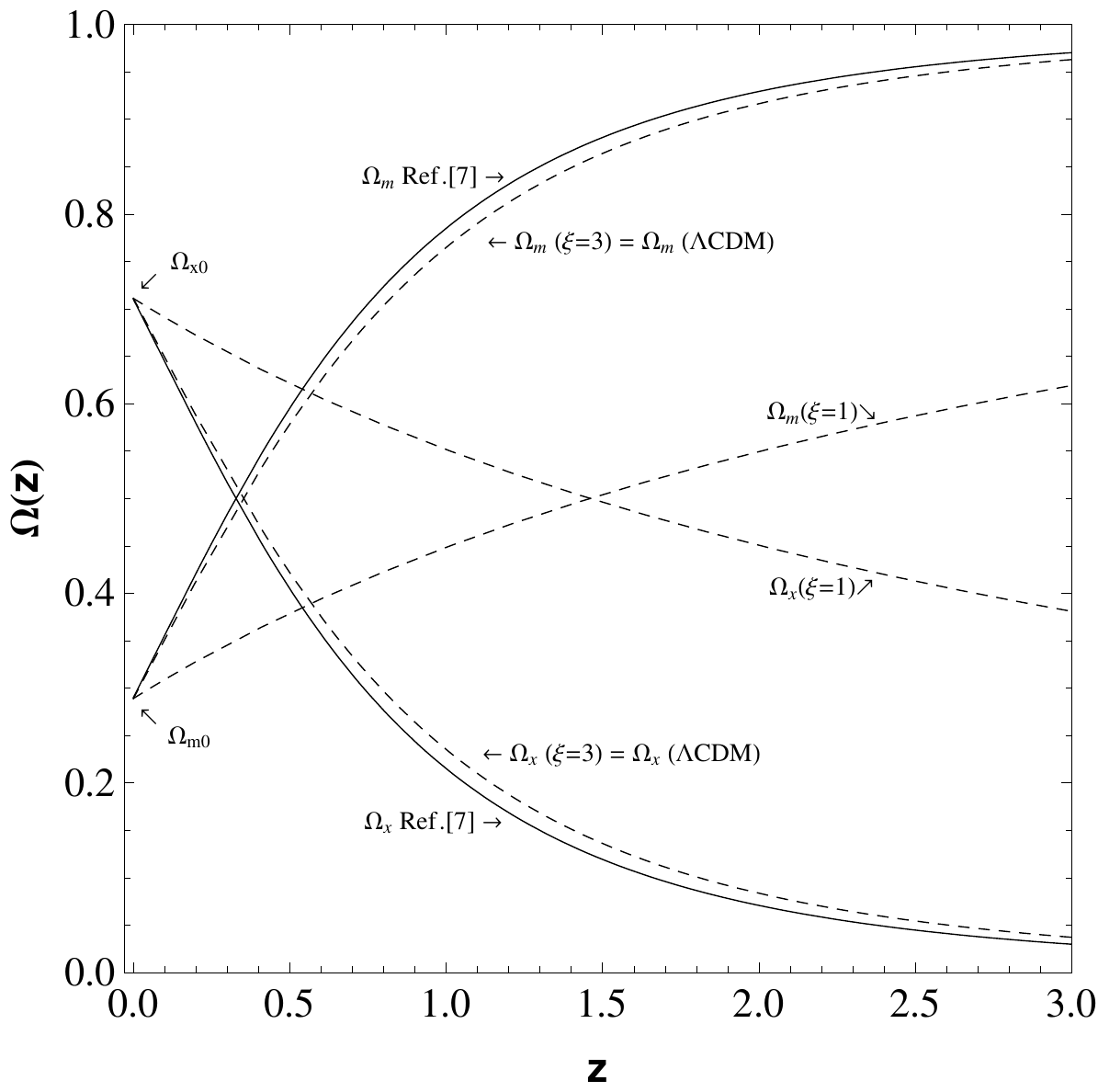}
\caption{Redshift dependence of the fractional contributions $\Omega_{m} = \frac{8\pi G \rho_{m}}{3 H^{2}}$ and $\Omega_{x} = \frac{8\pi G \rho_{x}}{3 H^{2}}$. The baryon contribution is neglected.
In this illustration the same values $\Omega_{m0}=0.289$ and $\omega = -1.01$ were used for all  the models.
While the curves for the interacting model of \cite{alcaniz} (solid lines) and for our $\xi=3$ model  (which is indistinguishable from the $\Lambda$CDM model) are similar to each other, the difference between $\Omega_{m}$ and $\Omega_{x}$ is much smaller for $\xi=1$ than for any of the other models.}
\label{figcoincidence}
\end{figure}

To illustrate the situation with respect to the coincidence problem, we compare the behavior of
$\Omega_{m} = 8\pi G \rho_{m}/(3 H^{2})$ and $\Omega_{x} = 8\pi G \rho_{x}/(3 H^{2})$
for the $\Lambda$CDM model, for the interacting model studied in Ref. \cite{alcaniz} (with a constant EoS parameter $\omega = -1.01$ and $\xi=3.16$) and for our interacting models with $\xi=1$ and $\xi=3$ in Fig.\ref{figcoincidence}. To plot and to compare the corresponding curves qualitatively, we rely on the fiducial best-fit values $\Omega_{m0}=0.289$ and $\omega = -1.01$ of Ref. \cite{alcaniz}, for all the models shown here.  For the case $\xi=1$, both curves are considerably closer to each other for a wide redshift range than for any of the other models, which we interpret as an alleviation of the coincidence problem. Fig. \ref{figcoincidence} shows also that for $\xi = 1$ the equality of both components occurs earlier in time than for the other models. In previous investigations some room was left in the parameter space for this model \cite{somasri}. We reconsider this issue here again on the basis of the recent data from SNIa, from BAO and from the CMB shift parameter.

\section{Statistical analysis and observational constraints}
\label{analysis}

As already mentioned, our aim is to compare models with the Hubble rates (\ref{Hbueno}),  (\ref{H1}) and (\ref{H3}).
To test the different models against cosmological observations, we consider the SNIa data sets Constitution \cite{Hicken} and Union2 \cite{Amanullah}, as well as baryon acoustic oscillations (BAO)
\cite{Eisenstein} and the CMB shift parameter \cite{komatsu}. For the SNIa data we adapt the procedure put forward in \cite{lazkoz}.

\subsection{Observational tests}

\subsubsection{SNIa}

Usually, the statistical $\chi^{2}$ analysis  procedure is based on the probability distribution ($\mathcal{N}$ is a normalization factor)
\begin{equation}\label{P(M)}
P(\overline{M},a_1,...,a_n)=\mathcal{N}e^{-\chi^{2}(\overline{M},a_1,...,a_n)},
\end{equation}
where $a_1,...,a_n$ is the relevant set of parameters and
\begin{equation}\label{m(z)}
m(z;a_1,...,a_n)=\overline{M}(M,H_0)+5\log_{10}(D_L(z;a_1,...,a_n))
\end{equation}
is the apparent magnitude of the SNIa which depends on the luminosity distance
\begin{equation}\label{dl1}
D_L^{th }(z;a_1,...,a_n)=
\left(1+z\right)\int_0^z \frac{\mbox{d}z^\prime}{H(z^\prime;a_1,...,a_n)}\ .
\end{equation}
The quantity $\overline{M}$ is related to the absolute magnitude by
\begin{equation}\label{M(z)}
\overline{M}=M+5\log_{10}\left(\frac{cH_0^{-1}}{Mpc}\right)+25.
\end{equation}
Moreover, for the cases of interest here, the Constitution dataset \cite{Hicken} with $397$
data and the $557$ data points from the Union2 dataset \cite{Amanullah}, we have
\begin{equation}\label{x2}
\chi^{2}(\overline{M},a_1,...,a_n)=\sum_{i=1}^{397/557}
\frac{(m^{obs}(z_i)-m^{th}(z_i;\overline{M},a_1,...,a_n)^2}{\sigma^2_{m^{obs}}(z_i)}\ ,
\end{equation}
where $\sigma_{m^{obs}}(z_i)$ denotes the $1\sigma$ errors of the SNIa data \cite{lazkoz}.
The parameters  $\overline{a}_1,...,\overline{a}_n$ that minimize the $\chi^2$ expression (\ref{x2}) are the most probable parameter values (the `best fit') and the corresponding  $\chi^2(\overline{a}_1,...,\overline{a}_n)\equiv \chi^2_{min}$ gives an indication of the quality of fit for the parametrization: the smaller $\chi^2_{min}$ is, the better the parametrization.
For our data analysis we follow the method developed in \cite{lazkoz} which implies a marginalization over $\overline{M}$ and adapt it to the Constitution and Union2 data sets. The crucial point consists in an expansion for
$\chi^2$ of equation (\ref{x2}) with respect to  $\overline{M}$ of the form
\begin{equation}\label{x2red}
\chi^{2}(a_1,...,a_n)=A-2\overline{M}B+\overline{M}^2C,
\end{equation}
where
\begin{eqnarray}\label{A11}
A(a_1,...,a_n)&=&\sum_{i=1}^{397/557}\frac{(m^{obs}(z_i)-m^{th}(z_i;\overline{M}=0,a_1,...,a_n)^2}{\sigma^2_{m^{obs}}(z_i)},\\
B(a_1,...,a_n)&=&\sum_{i=1}^{397/557}\frac{(m^{obs}(z_i)-m^{th}(z_i;\overline{M}=0,a_1,...,a_n)}{\sigma^2_{m^{obs}}(z_i)},\\
C&=&\sum_{i=1}^{397/557}\frac{1}{\sigma^2_{m^{obs}}(z_i)}.
\end{eqnarray}
Equation (\ref{x2red}) has a minimum  for $\overline{M}=B/C$ at
\begin{equation}\label{x222}
\tilde{\chi}^{2}(a_1,...,a_n)=A(a_1,...,a_n)-\frac{B(a_1,...,a_n)^2}{C} .
\end{equation}
Thus, instead of minimizing $\chi^{2}(\overline{M},a_1,...,a_n)$ we can minimize $\tilde{\chi}^{2}(a_1,...,a_n)$ which is independent of $\overline{M}$. Obviously,  $\chi^{2}_{min}=\tilde{\chi}^{2}_{min}$.

\begin{figure}[!h]
\centering
\includegraphics[width=0.60\textwidth]{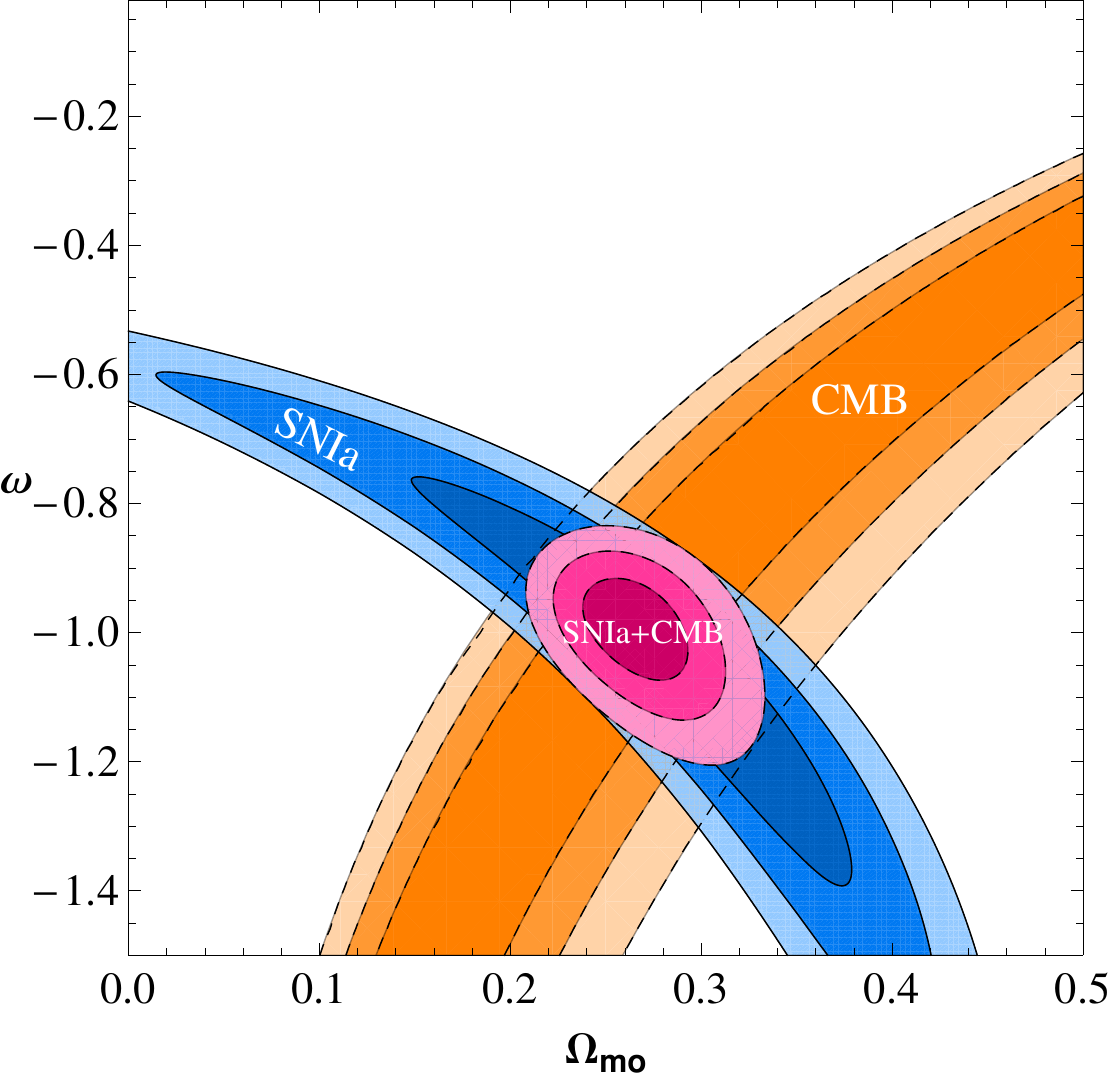}
\caption{Reproduction (for SNIa+CMB) of the left panel of Fig.~12 in \cite{Amanullah} for the $\omega$CDM model with two free parameters for SNIa+CMB. Our best-fit values (68.3\% confidence level) are $\Omega_{m0} = 0.266^{+0.018}_{-0.017}$ and $\omega = -0.994^{+0.048}_{-0.052}$ compared with $\Omega_{m0} = 0.268^{+0.019}_{-0.017}$ and $\omega = -0.997^{+0.050}_{-0.055}$ in \cite{Amanullah}.}
\label{figAm}
\end{figure}

To be more specific, for the numerical analysis we use the data of table 1 (SALT) in \cite{Hicken} and of the Union2 data set \cite{Amanullahdata}. To gauge our code, we reproduced in Fig.~\ref{figAm}    the left panel of Figure 12 of \cite{Amanullah} (SNIa + CMB) for a $\omega$CDM model  with a constant EoS parameter $\omega$ without systematic errors and generalized it to the case of time-dependent EoS parameters.

\subsubsection{BAO}

Acoustic oscillations at recombination give rise to a peak in the large-scale correlation function in luminous red galaxies
\cite{Eisenstein,Percival,Nesseris-Perilovorapoulos}.
This peak can be related to a distance scale
\begin{equation}
D_{v}(z_{BAO})=\left[\frac{z_{BAO}}{H(z_{BAO})}
\left(\int^{z_{BAO}}_{0}\frac{dz}{H(z)}\right)^{2}\right]^{\frac{1}{3}}
\ ,
\label{Dv}
\end{equation}
characterized by a dimensionless parameter
\begin{equation}
A(z_{BAO};a_{1}...a_{n})=\sqrt{\Omega_{m0}}E(z_{BAO})^{-1/3}
\left[\frac{1}{z_{BAO}}\int_0^{z_{BAO}}\frac{dz}{E(z;a_{1}...a_{n})}\right]^{2/3}\
\end{equation}
with $z_{BAO}=0.35$ and $E(z) \equiv \frac{H(z)}{H_{0}}$.
The observational value is $A=0.469 \pm 0.017$ \cite{Eisenstein}. For $\chi^2_{BAO}$  we have
\begin{equation}
\chi^2_{BAO}(a_{1}...a_{n})=\frac{(A(z_{BAO};a_{1}...a_{n})-0.469)^2}{0.017^2}\ .
\label{xb}
\end{equation}

\subsubsection{CMB}

The CMB shift parameter measures the displacement of the first acoustic peak of the CMB anisotropy spectrum with respect to the position this peak would have in a flat Einstein-de Sitter reference universe.
For a flat universe it is given by \cite{wang-mukherjee,bond}
\begin{equation}
R(z_{ls};a_{1}...a_{n})=\sqrt{\Omega_m}\int_0^{z_{ls}}\frac{dz}{E(z;a_{1}...a_{n})}
\end{equation}
with the last scattering redshift $z_{ls}=1090$.
The 7-years WMAP  result for this parameter is $R=1.725 \pm 0.018$ \cite{komatsu}. For the $\chi^2_{CMB}$ value we have
\begin{equation}
\chi^2_{CMB}=\frac{(R-1.725)^2}{0.018^2} .
\label{xc}
\end{equation}

For our analysis we combine the $\chi^2$ values from the three tests (\ref{x222}), (\ref{xb}) and (\ref{xc}) to minimize the total $ \chi^2_{total} $
\begin{equation}
\chi^2_{total}=\tilde{\chi}^{2}_{SNIa}+\chi^2_{BAO}+\chi^2_{CMB}\ .
\end{equation}

\subsection{Analysis of the  models}

For a non-interacting two-component model of the dark sector the consequences of a time-variable EoS parameter, based on the CPL parametrization and the Constitution data set, have been studied in \cite{bueno}. For the Union2 set a similar analysis was performed in \cite{LiWuYu}.
Our goal here is to extend this type of analysis to cosmological models with interactions in the dark sector.
We shall rely on the CPL parametrization for the dark-energy EoS as well and use both the Constitution and the Union2 data sets.

Our interacting models have 4 free parameters, namely the Hubble constant $H_0$, the fractional density of dark matter $\Omega_{m0}$ and the EoS parameters  $\omega_0$ and $\omega_1$. The baryon fraction $\Omega_{b0}$ is assumed to have the fixed value $\Omega_{b0} = 0.042$.
As usual, $H_{0}$ is parametrized by $H_{0} = \mathrm{100}\ h\ \mathrm{\mathrm{km\ sec}^{-1}\mathrm{Mpc}^{-1}}$. The parameter $h$ will be marginalized according to (\ref{x222}). Hence, our models are left with the 3 free parameters  $\Omega_{m0}$, $\omega_0$ and $\omega_1$.

If baryons are taken into account, only the quantity $\Omega_{dm0}=\Omega_{m0}-0.042$ interacts since baryons (with $\Omega_{b0}=0.042$) are separately conserved. Generally, we find that the inclusion of baryons has only a very small influence on the curves. Therefore, baryons do not appear explicitly in our graphic representations.

\subsubsection{The $\Lambda$CDM model}

For later comparison we start with the best-fit data for the $\Lambda$CDM model in tables~\ref{table1lcdm} and \ref{table2lcdm}. These results are consistent with those of \cite{NessDeFeliceTsuji}.
\begin{table}[ht]
\centering
\begin{tabular}{l  l  l  l} 
\hline\hline                        
\makebox [4 cm ][l]{Observation} &\makebox [2 cm ][l]{$\chi^2_{min}$} & \makebox [2 cm ][l]
{$\Omega_{m0}$} & \makebox [2 cm ][l]{$q_0$}   \\ [0.5ex] 
\hline\hline                   
SNIa & 465.513 &0.289 &-0.566\\ 
SNIa+BAO&465.731& 0.282  &-0.577\\
SNIa+CMB & 466.179& 0.278 &-0.583\\
SNIa+BAO+CMB & 466.202 & 0.276 &-0.585\\ [1ex]      
\hline\hline  
\end{tabular}
\caption{Best-fit values based on the Constitution set for the $\Lambda$CDM model.}
\label{table1lcdm}
\end{table}

\begin{table}[ht]
\centering
\begin{tabular}{l  l  l  l} 
\hline\hline                        
\makebox [4 cm ][l]{Observation} &\makebox [2 cm ][l]{$\chi^2_{min}$} & \makebox [2 cm ][l]
{$\Omega_{m0}$}  & \makebox [2 cm ][l]{$q_0$}  \\ [0.5ex] 
\hline\hline                   
SNIa & 541.012 & 0.269 &-0.596\\ 
SNIa+BAO& 541.029& 0.271 &-0.593 \\
SNIa+CMB & 541.091& 0.266 &-0.601 \\
SNIa+BAO+CMB &541.156 & 0.268 &-0.598\\ [1ex]      
\hline\hline  
\end{tabular}
\caption{Best-fit values based on the Union2 set for the $\Lambda$CDM model.}
\label{table2lcdm}
\end{table}

\subsubsection{The non-interacting case}

\begin{figure}[!h]
\centering
\includegraphics[width=0.402\textwidth]{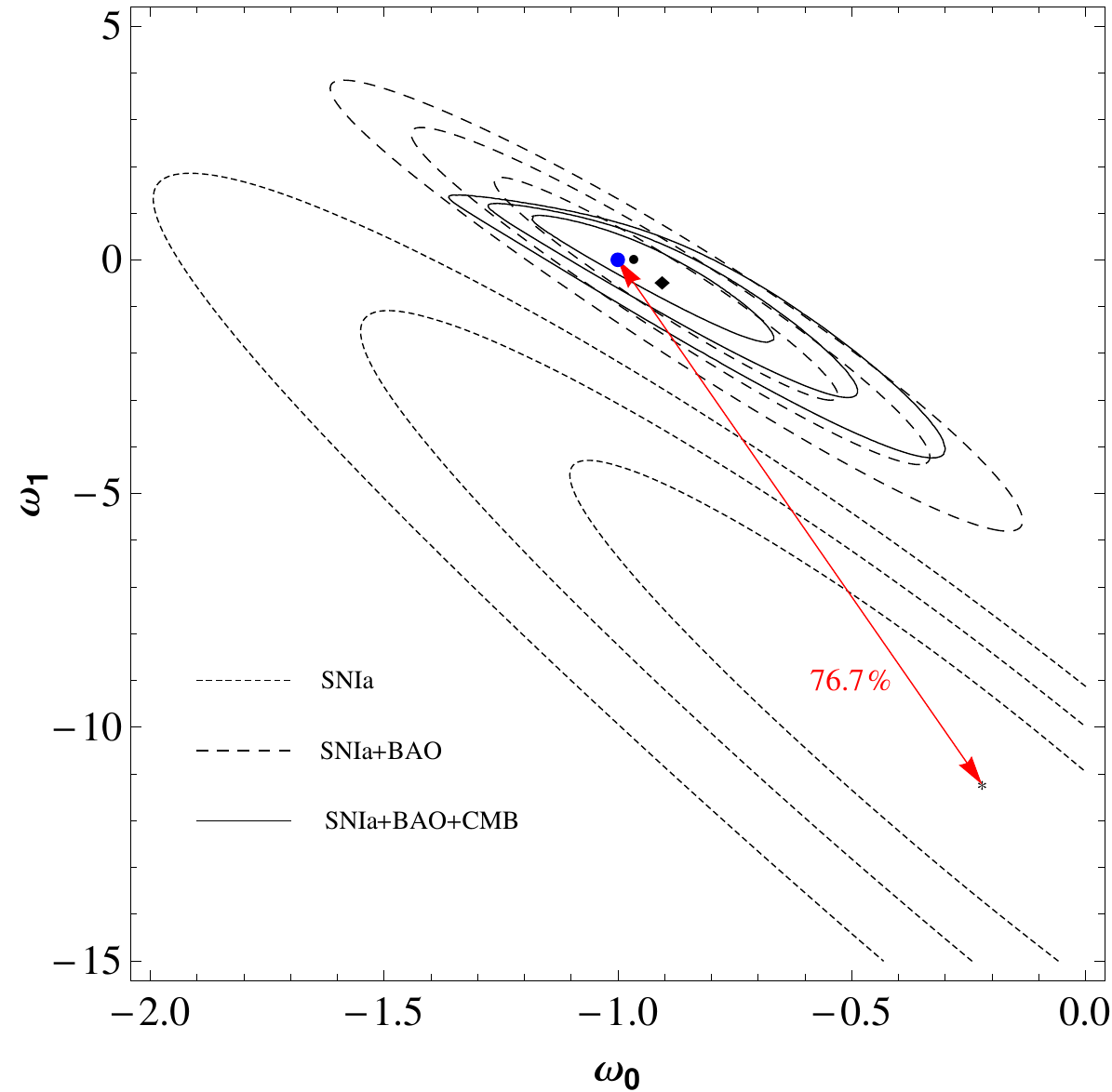}	
\includegraphics[width=0.4\textwidth]{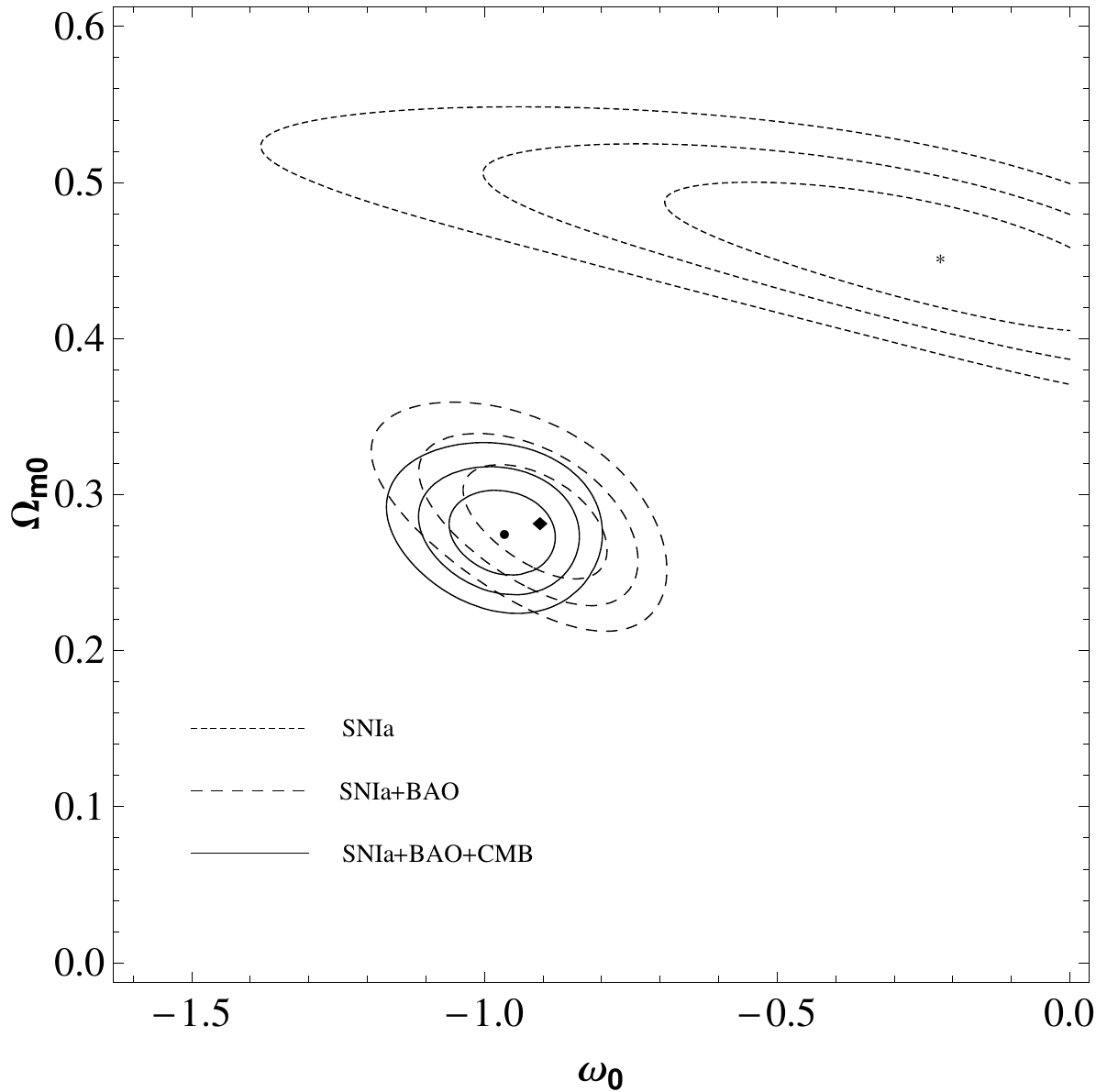}	
\caption{Non-interacting case: Contour plots ($1\sigma$, $2\sigma$ and $3\sigma$) in the $\omega_{0}$-$\omega_{1}$ plane (left panel) with delta priors for $\Omega_{m0}$ from Table \ref{tableni1}. Contour plots in the $\omega_{0}$-$\Omega_{m0}$ plane (right panel) with corresponding delta priors for $\omega_1$.
The red arrow in the left panel characterizes the difference $\Delta\chi^2 =\chi^2(\Lambda\mathrm{CDM})-\chi^2 (\mathrm{Case\ No\ Int.})$ for the case of three free parameters
(cf. Table~\ref{tableP}).}
\label{figni1}
\end{figure}

In a next step we consider the non-interacting case based on (\ref{Hbueno}).
The confidence level  contours ($1\sigma$, $2\sigma$ and $3\sigma$) for the Constitution set (short-dashed lines for SN only, long-dashed for a joint analysis SN+BAO and solid lines for the joint analysis SN+BAO+CMB) are shown in Fig.~\ref{figni1}.
 The 2D contours are produced on the basis of 3 free parameters, i.e. $\Delta \chi^2 \equiv \chi^2 - \chi^2_{min} \leq 3.53 (68\%), 8.02 (95\%)$ and $14.02 (99.73\%)$. To obtain the contours in the $\omega_{0}$ - $\omega_{1}$ plane (left panel), delta priors for $\Omega_{m0}$  according to the best-fit values in Table \ref{tableni1} were used.
For the SN-only case, e.g., this means $\Omega_{m0}=0.4519$. The contours in the right panel rely on corresponding delta priors for $\omega_1$, i.e., $\omega_1=-11.227$ for the SN-only case etc. These curves reproduce the results of \cite{haowei1}. The corresponding curves for the Union2 sample are similar.
The red arrow in the left panel characterizes the distance between the non-interacting model and the $\Lambda$CDM model (blue dot) for three free parameters if only the SNIa data are used. Under this condition, a value of  $\Delta\chi_{min}^2 = 4.28$ (cf. Table~\ref{tableP}) corresponds to the $76.7\%$ confidence region, equivalent to $1.31 \sigma$. Table~\ref{tableP} visualizes the relation between  $\Delta \chi^2$ and the joint posterior probability $P$, given by \cite{gregory}
\begin{equation}\label{P}
P = 1 - \frac{\gamma\left(\nu/2, \Delta\chi^{2}/2\right)}{\Gamma(\nu/2)}\ ,
\end{equation}
where $\nu$ is the number of free parameters and $\gamma\left(\nu/2, \Delta\chi^{2}/2\right)$ is the
\textit{incomplete gamma function}.

\begin{figure}[!h]
\centering
\includegraphics[width=0.52\textwidth]{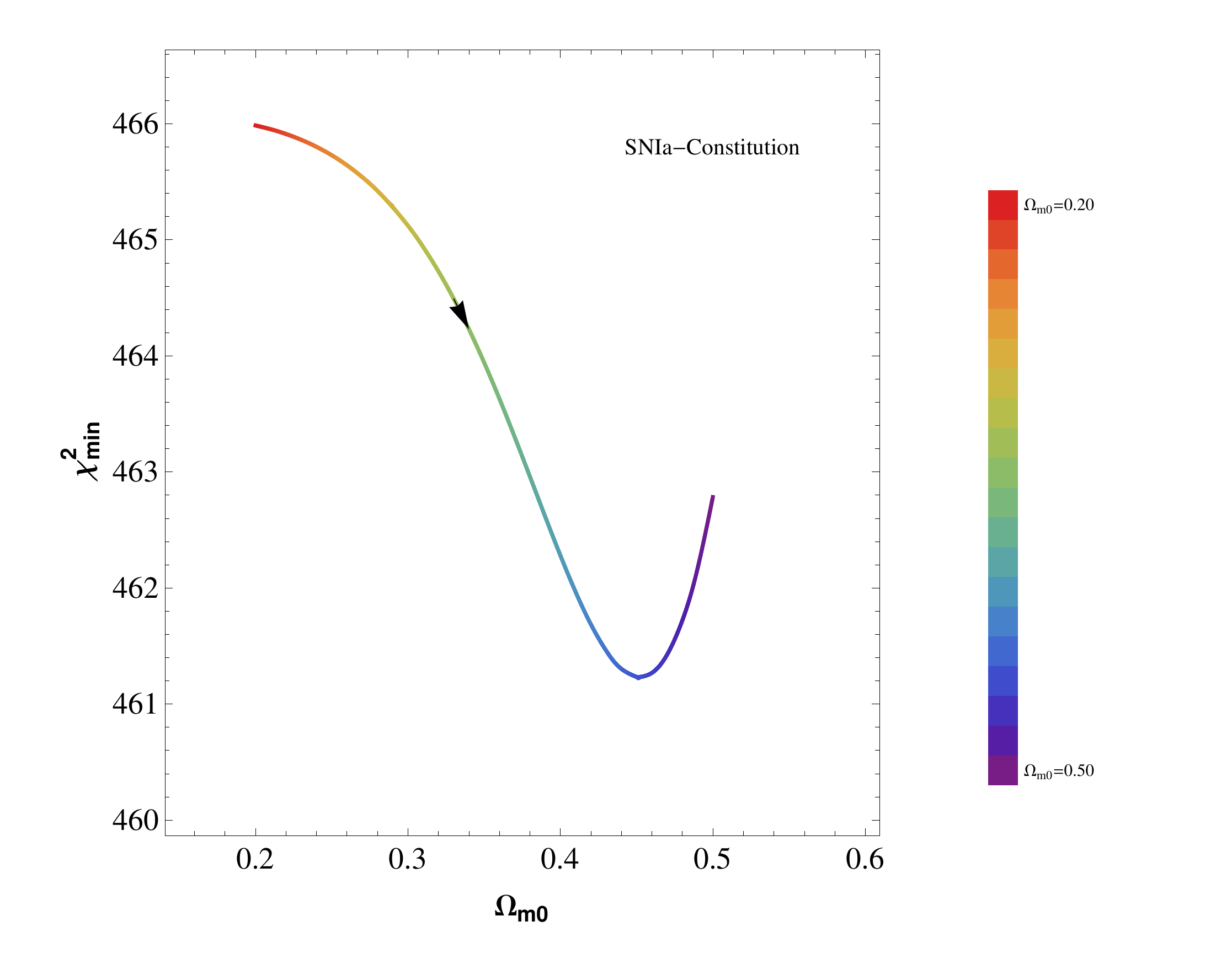}	
\includegraphics[width=0.38\textwidth]{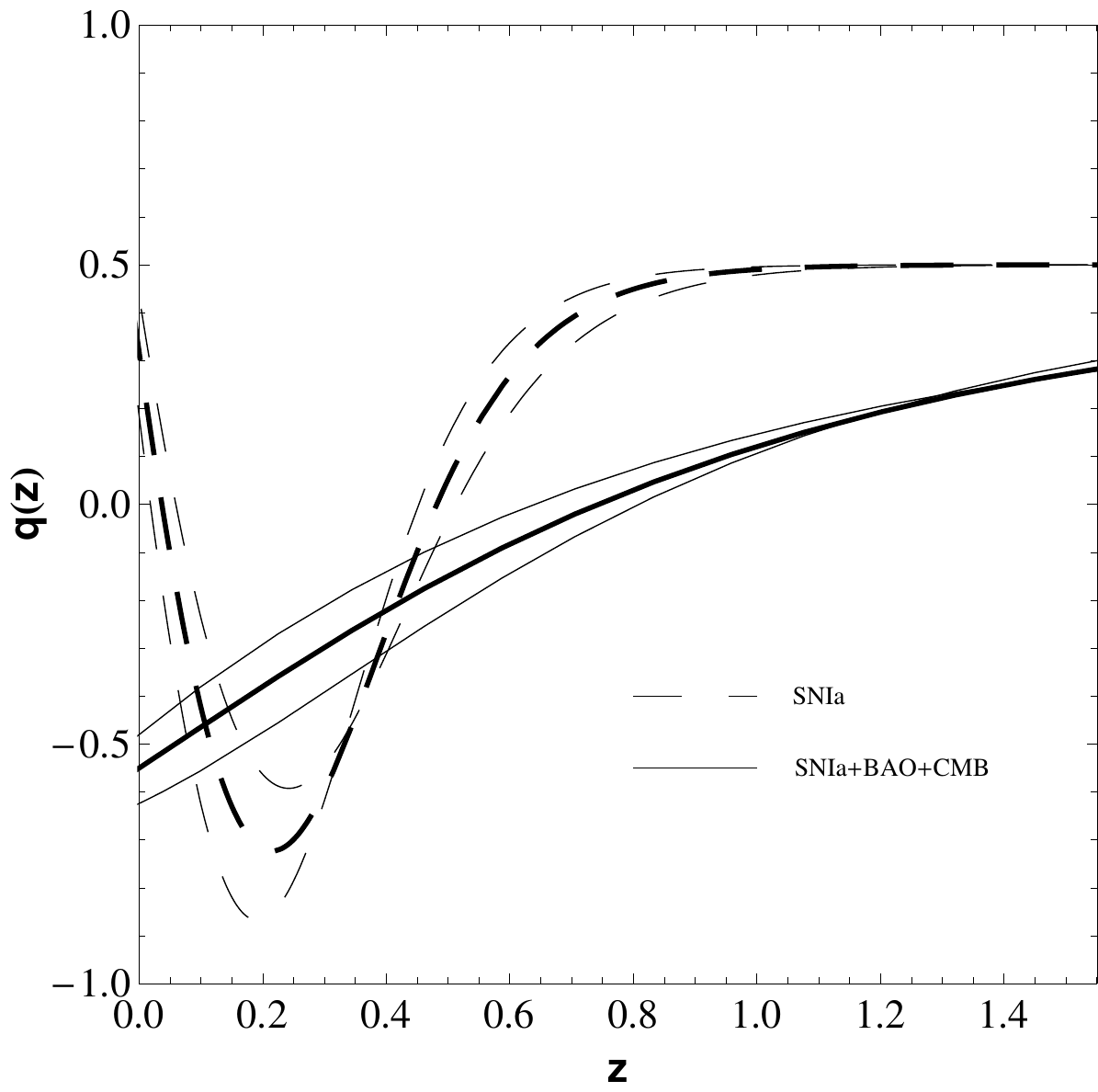}	
\caption{Left panel: Dependence of $\chi^{2}_{min}$ on $\Omega_{m0}$ for the Constitution set. Right panel: Dependence of the deceleration parameter on the redshift (with $1\sigma$ contour lines) for two  data-set combinations. For SNIa only, the best-fit model ($\Omega_{m0}=0.4519$, $\omega_0 = -0.221$, $\omega_1 = - 11.227$ (cf.~Table \ref{tableni1})) predicts a present decelerated expansion $q_{0} >0$.}
\label{figni2}
\end{figure}

The left panel of Fig.~\ref{figni2} shows the dependence of $\chi^{2}_{min}$ on $\Omega_{m0}$
for the range $0.20 < \Omega_{m0} < 0.50$ if only SNIa data (here, the Constitution set) are used.
The minima correspond to much larger  values of $\Omega_{m0}$ ($0.452$ for Constitution and $0.415$ for Union2) than for the $\Lambda$CDM model ($0.289$ for Constitution and $0.269$ for Union2). If combined, however, with BAO and CMB, these values reduce substantially, as shown in Table~\ref{tableni1} (Constitution) and Table~\ref{tableni2} (Union2).

\begin{table}[ht]
\centering
\begin{tabular}{l  l  l  l  l  l  l} 
\hline\hline                        
\makebox [4 cm ][l]{Observation} &\makebox [2 cm ][l]{$\chi^2_{min}$} & \makebox [2 cm ][l]{$\Omega_{m0}$} & \makebox [2 cm ][l]{$\omega_0$} & \makebox [2 cm ][l]{$\omega_1$} & \makebox [2 cm ][l]{$q_0$}  \\ [0.5ex] 
\hline\hline                   
SNIa &   461.231  &   0.452 &   -0.221 &  -11.227  &  0.318\\ 
SNIa+BAO & 465.425 &0.281 & -0.905& -0.497  & -0.475\\
SNIa+CMB &465.557 & 0.270  & -0.935 & -0.119 & -0.524\\
SNIa+BAO+CMB & 465.606 & 0.274 & -0.965& 0.015 & -0.550\\ [1ex]      
\hline\hline  
\end{tabular}
\caption{Best-fit values based on the Constitution set for the non-interacting model with the Hubble rate (\ref{Hbueno}).}
\label{tableni1} 
\end{table}

\begin{table}[ht]
\centering
\begin{tabular}{l  l  l  l  l  l  l} 
\hline\hline                        
\makebox [4 cm ][l]{Observation} &\makebox [2 cm ][l]{$\chi^2_{min}$} & \makebox [2 cm ][l]
{$\Omega_{m0}$} & \makebox [2 cm ][l]{$\omega_0$} & \makebox [2 cm ][l]{$\omega_1$} & \makebox [2 cm ][l]{$q_0$} \\ [0.5ex] 
\hline\hline                   
SNIa &539.878 & 0.415 &  -0.886 & -5.108 &-0.278 \\ 
SNIa+BAO &540.988& 0.274 &-1.007& -0.039&-0.598\\
SNIa+CMB &541.028& 0.266&-1.010& 0.103&-0.612\\
SNIa+BAO+CMB &541.070& 0.269&-1.030& 0.187&-0.630\\ [1ex]      
\hline\hline  
\end{tabular}
\caption{Best-fit values based on the Union2 set for the non-interacting model with the Hubble rate  (\ref{Hbueno}).}
\label{tableni2}
\end{table}
These tables also contain the best-fit values for $\omega_{0}$ and $\omega_{1}$ for different data-set combinations and the corresponding values for the deceleration parameter. The right panel of Fig.~\ref{figni2} displays the reconstruction of the deceleration parameter from the data. Note that using only the SNIa data from the Constitution set, the best-fit model  ($\Omega_{m0}=0.4519$, $\omega_0 = -0.221$, $\omega_1 = - 11.227$ (cf.~Table \ref{tableni1})) predicts a decelerated expansion $q_{0} >0$ (cf. \cite{sastaro}). This is different from the result $q_{0} <0$ based on the Union2 sample (cf. Table~\ref{tableni2})
All the results for the non-interacting model, based on the Constitution set, are consistent with those of \cite{bueno} and \cite{haowei1}, where the latter reference also discusses the tensions between the different SNIa data sets.

\begin{table}[ht]
\centering
\begin{tabular}{l  l  l  l  l  l } 
\hline\hline                        
\makebox [4 cm ][l]{Observation} &No Int. & $\Lambda$CDM & \makebox [2 cm ][l]{$\Delta\chi^2$} &\makebox [2 cm ][l] {$\quad P$} \\ [0.5ex] 
\hline\hline                   
SNIa &   461.231  &   465.513 &  4.282 &  76.7\% (1.31$\sigma$) \\ 
SNIa+BAO & 465.425 &465.731 & 0.306& 4.1\% ($ \simeq $0.06$\sigma$) \\
SNIa+CMB &465.557 &466.179  & 0.622 & 10.9\% ($ \simeq $0.16$\sigma$) \\
SNIa+BAO+CMB & 465.606 &466.202 & 0.596& 10.3\% ($ \simeq $0.15$\sigma$) \\ [1ex]
\hline\hline  
\end{tabular}
\caption{Minimum $\chi^2$ values for the non-interacting and the $\Lambda$CDM models and the differences  $\Delta\chi^2 =\chi^2(\Lambda\mathrm{CDM})-\chi^2 (\mathrm{Case\ No\ Int.})$. The right column provides the corresponding probability values for the case of three free parameters. For comparison we recall that $\Delta\chi^2=3.53$ corresponds to
$P = 68.3\% (1\sigma)$, $\Delta\chi^2=8.02$ corresponds to
$P = 95.4\% (2\sigma)$ and $\Delta\chi^2=14.2$ corresponds to
$P = 99.73\% (3\sigma)$ \cite{gregory}.}
\label{tableP}
\end{table}

\subsubsection{The case $\xi=1$}

\begin{figure}[!h]
\centering
\includegraphics[width=0.4\textwidth]{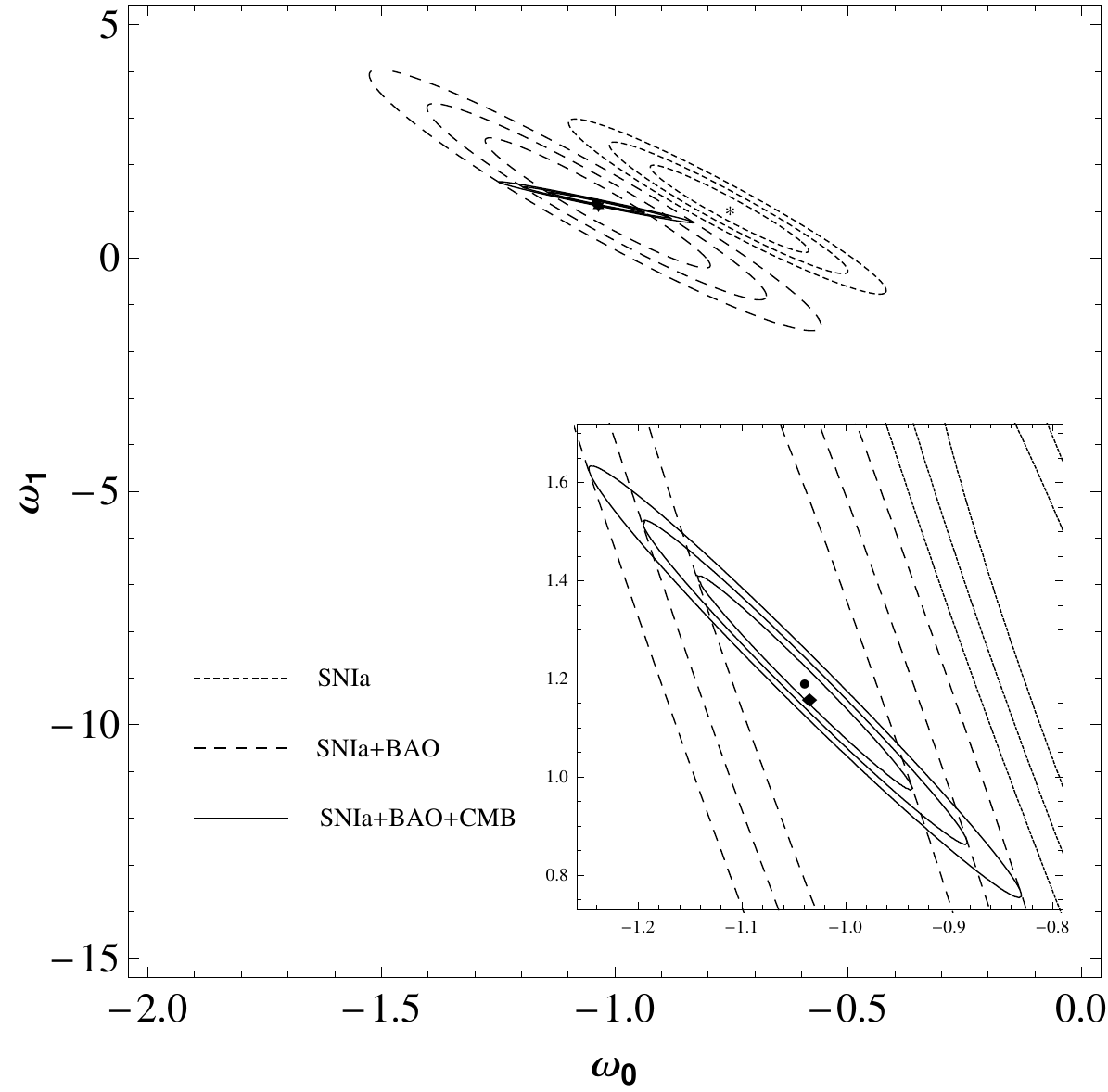}	
\includegraphics[width=0.397\textwidth]{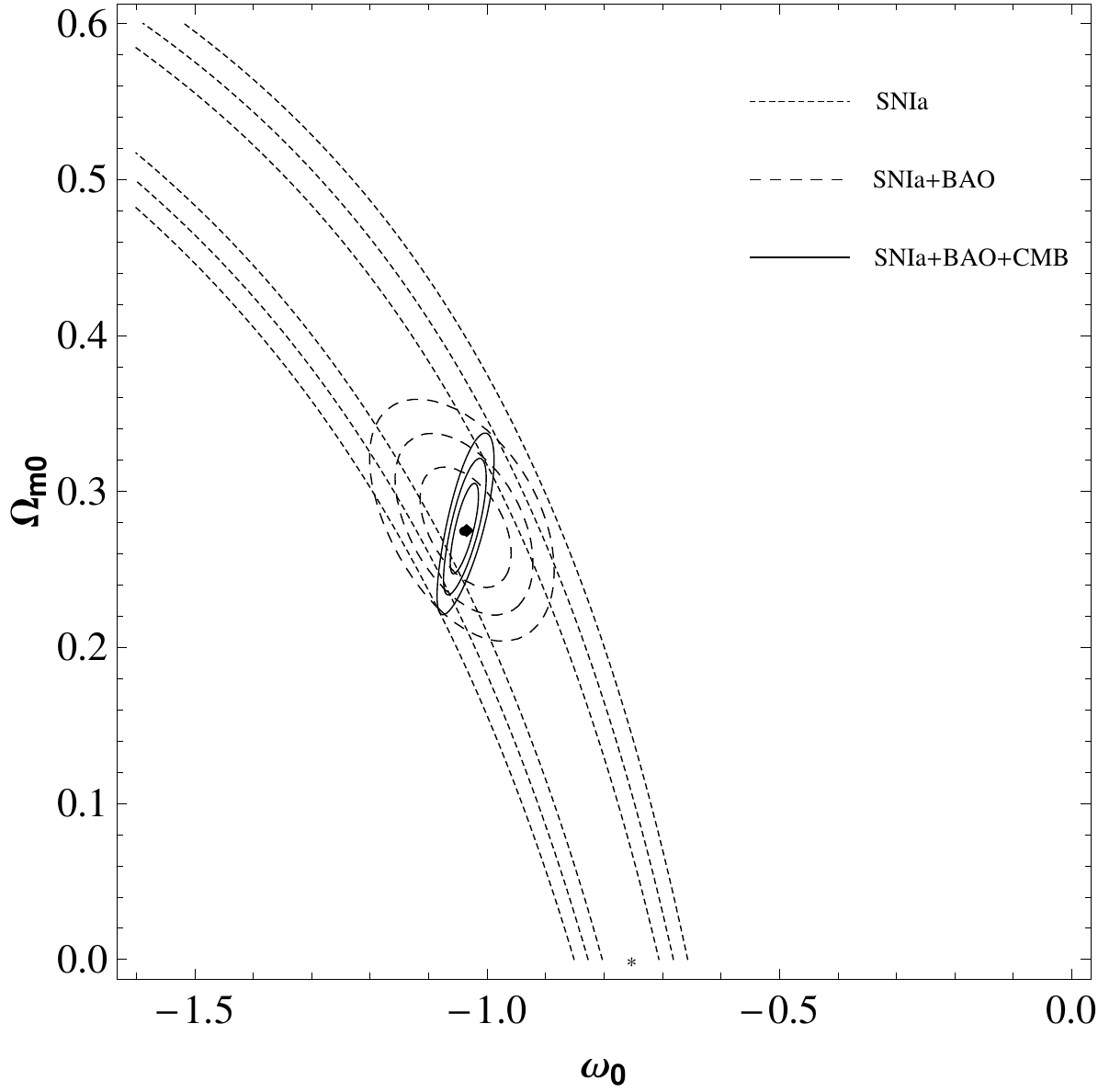}	
\caption{Case $\xi=1$: Contour plots ($1\sigma$, $2\sigma$ and $3\sigma$) in the $\omega_{0}$-$\omega_{1}$ plane (left panel) with delta priors for $\Omega_{m0}$ from Table \ref{table1x1}. Contour plots in the $\omega_{0}$-$\Omega_{m0}$ plane (right panel) with delta priors for $\omega_{1}$ from Table \ref{table1x1}.}
\label{fig1x1}
\end{figure}

\begin{figure}[!h]
\centering
\includegraphics[width=0.5\textwidth]{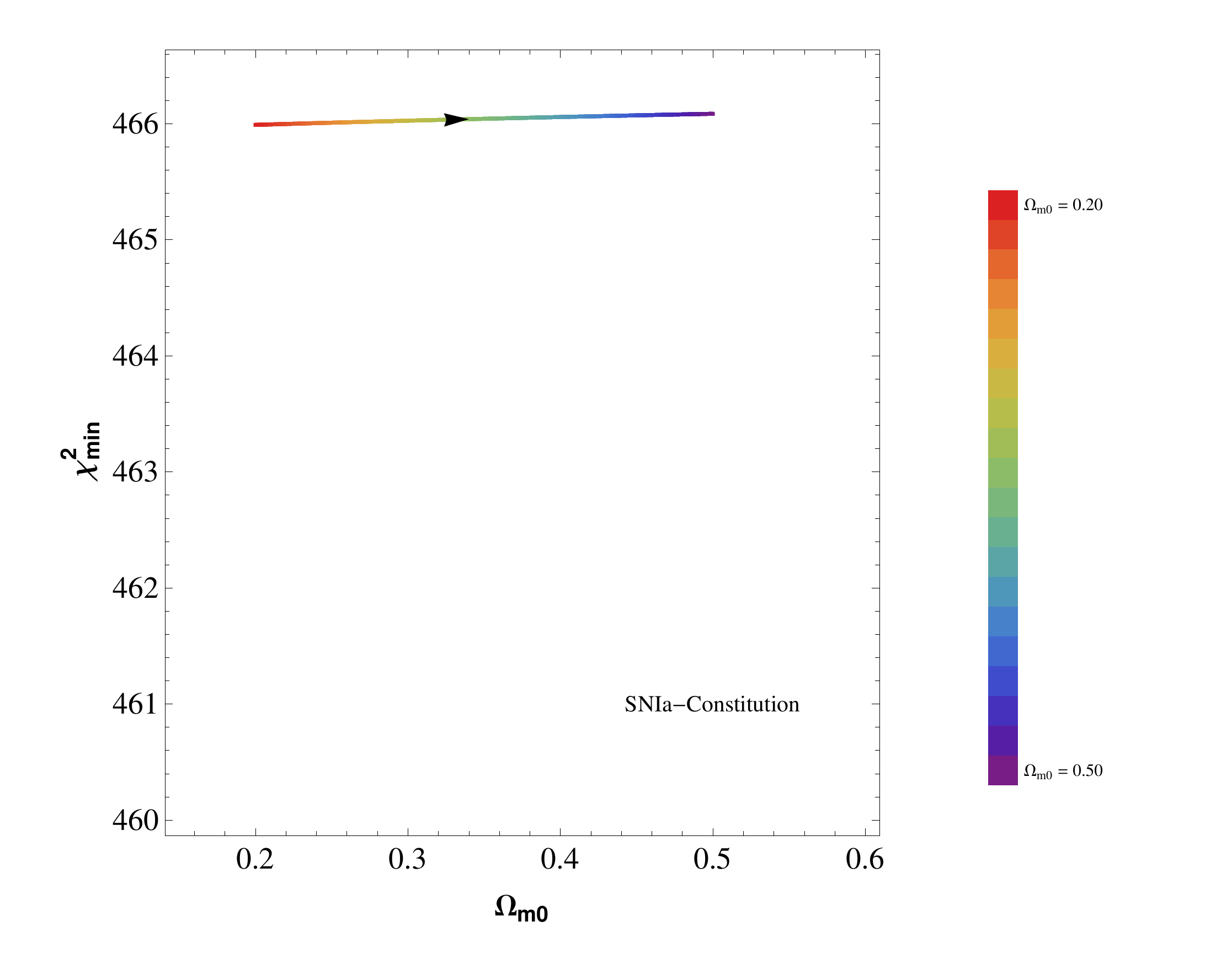}	
\includegraphics[width=0.38\textwidth]{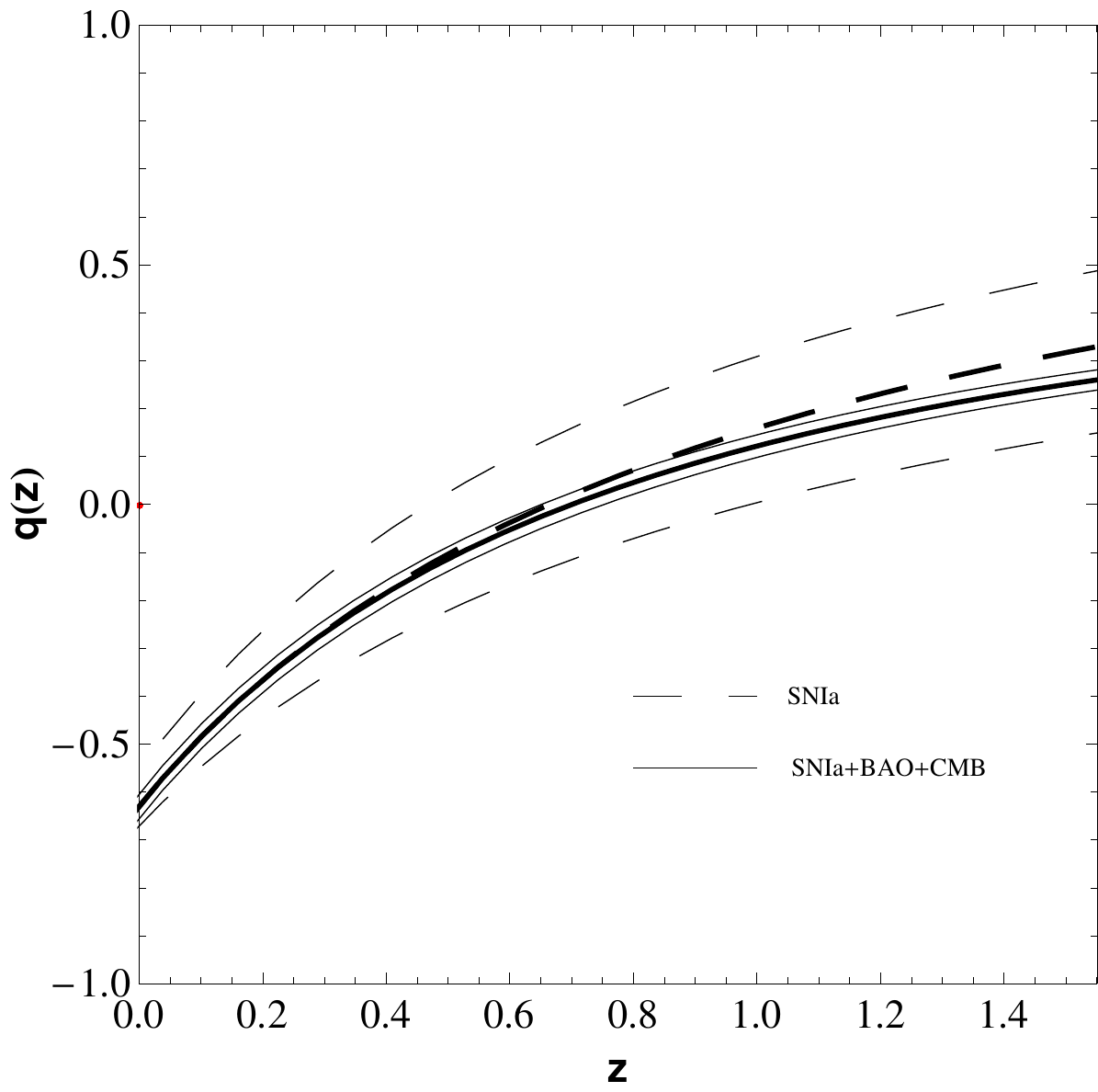}	
\caption{Case $\xi=1$. Left panel: dependence of $\chi^{2}_{min}$ on $\Omega_{m0}$.
 Notice that $\chi^{2}_{min}$ remains practically constant over the entire range of $\Omega_{m0}$. Right panel: dependence of the deceleration parameter on the redshift (with $1\sigma$ contour lines) for two data-set combinations.}
\label{fig2x1}
\end{figure}

The results for this model are shown in Figs.~\ref{fig1x1} and \ref{fig2x1}, the corresponding best-fit values are given in tables \ref{table1x1} and \ref{table2x1}.
Different from the non-interacting case, the minimum value of $\chi^{2}$ for the SNIa samples is practically independent of $\Omega_{m0}$. But there is a slight tendency  to $\Omega_{m0}\sim 0$ (left panel of Fig.~\ref{fig2x1}). This degeneracy is also seen in the solid lines of the right panel of Fig~.~\ref{fig1x1} where only SN data (Constitution) have been used. It shows elongated (with respect to $\Omega_{m0}$) contour plots and has more negative values for $\omega_{0}$ than the non-interacting case.
For the SNIa analysis we included in tables \ref{table1x1} and \ref{table2x1} also the results for a prior $\Omega_{m0} = 0.289$.
There is a tendency to larger values of $\omega_{1}$ compared with the non-interacting case. The left panel of Fig.~\ref{fig1x1} reveals that both data set combinations indicate $\omega_{1} \sim 1$ while $\omega_{0} \sim -1$ for the joint analysis, something different from the non-interacting case as well. The insert magnifies the relevant region. The deceleration parameter behaves almost in the same way for the different data combination as can be seen in the right panel of Fig.~\ref{fig2x1}.

 \begin{table}[ht]
\centering  
\begin{tabular}{l  l  l  l  l  l  l} 
\hline\hline                        
\makebox [4 cm ][l]{Observation} &\makebox [2 cm ][l]{$\chi^2_{min}$} & \makebox [2 cm ][l]
{$\Omega_{m0}$} & \makebox [2 cm ][l]{$\omega_0$} & \makebox [2 cm ][l]{$\omega_1$} & \makebox [2 cm ][l]{$q_0$} \\ [0.5ex] 
\hline\hline                   
SNIa &   465.906  &  0.000 & -0.753 & 1.051 &-0.629 \\
(Prior $\Omega_{m0}=0.289$)& 466.022    &  0.289 & -1.054 &1.167  &-0.624 \\
\hline
SNIa+BAO & 466.018 &0.276 & -1.035 & 1.159 &-0.624\\
SNIa+CMB &  466.015&  0.266 & -1.018 &1.135&-0.621 \\
SNIa+BAO+CMB & 466.020 & 0.275 & -1.039& 1.191&-0.630 \\ [1ex]      
\hline\hline  
\end{tabular}
\caption{Best-fit values based on the Constitution set for the case $\xi=1$ with the Hubble rate  (\ref{H1}).}
\label{table1x1}
\end{table}

\begin{table}[ht]
\centering
\begin{tabular}{l  l  l  l  l  l  l} 
\hline\hline                        
\makebox [4 cm ][l]{Observation} &\makebox [2 cm ][l]{$\chi^2_{min}$} & \makebox [2 cm ][l]
{$\Omega_{m0}$} & \makebox [2 cm ][l]{$\omega_0$} & \makebox [2 cm ][l]{$\omega_1$} & \makebox [2 cm ][l]{$q_0$} \\ [0.5ex] 
\hline\hline                   
SNIa &   541.223  &   0.000 & -0.779 & 1.045 &-0.669\\
(Prior $\Omega_{m0}=0.269$)& 541.275    &  0.269 & -1.063 &1.171  &-0.665 \\
\hline
SNIa+BAO & 541.275 &0.271 & -1.066 & 1.173 &-0.665\\
SNIa+CMB & 541.272 & 0.252  & -1.037 & 1.153&-0.664\\
SNIa+BAO+CMB & 541.300 & 0.270 & -1.081& 1.269 &-0.683\\ [1ex]      
\hline\hline  
\end{tabular}
\caption{Best-fit values based on the Union2 set for the case $\xi=1$ with the Hubble rate (\ref{H1}).}
\label{table2x1}
\end{table}

\subsubsection{The case $\xi=3$}
\begin{figure}[!h]
\centering
\includegraphics[width=0.4\textwidth]{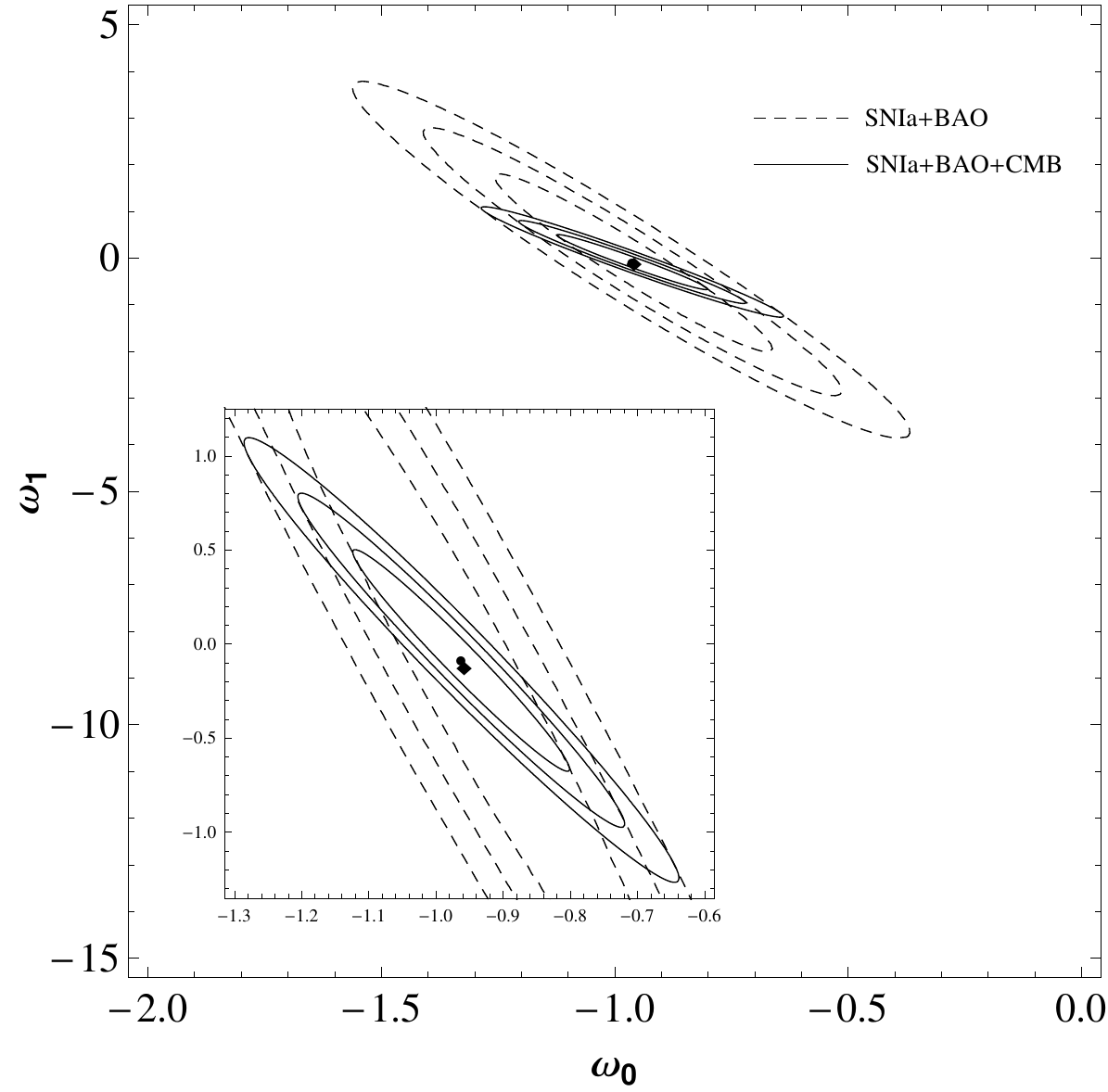}	
\includegraphics[width=0.415\textwidth]{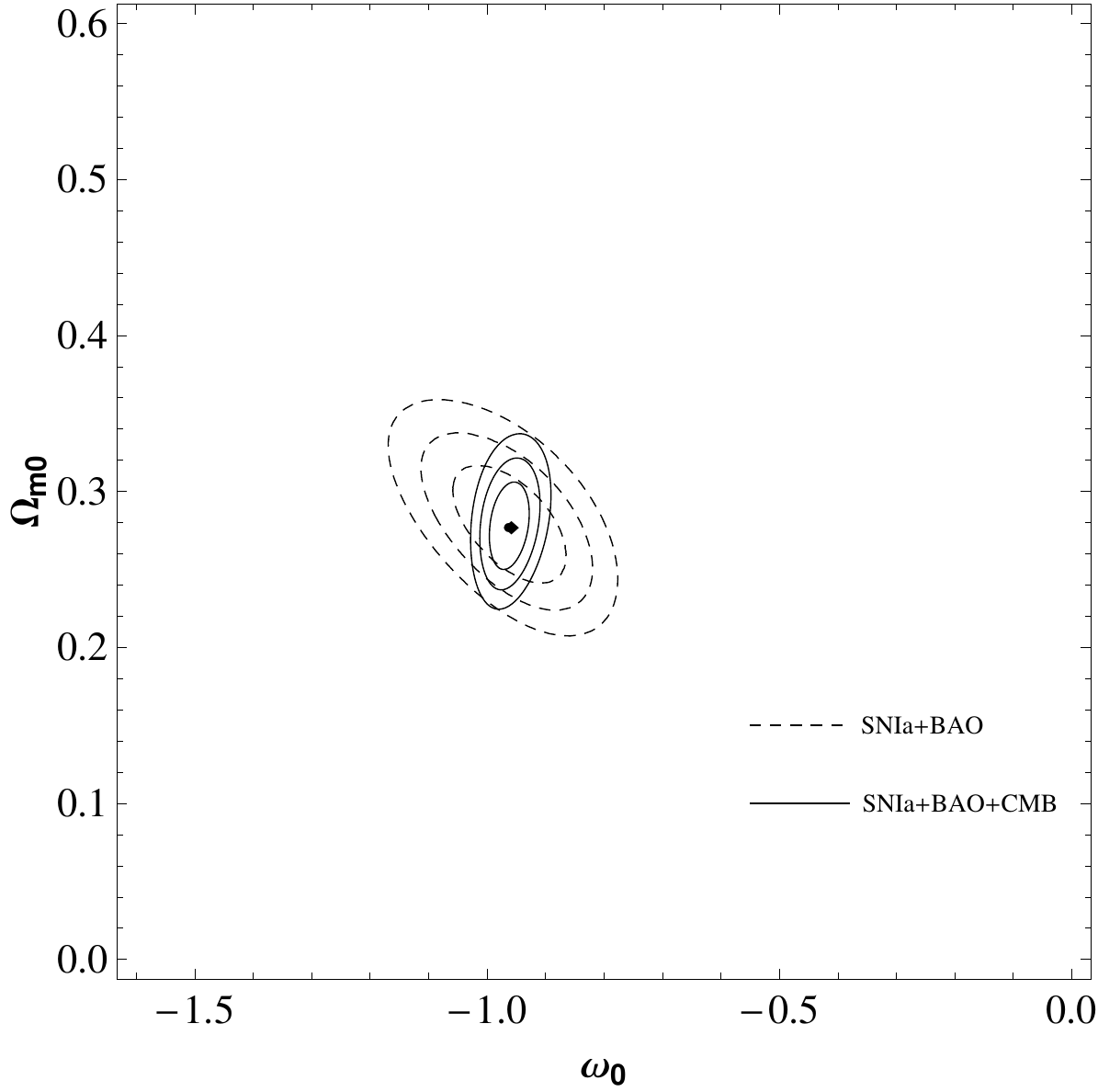}	
\caption{Case $\xi=3$: Contour plots ($1\sigma$, $2\sigma$ and $3\sigma$) in the $\omega_{0}$-$\omega_{1}$ plane (left panel) with delta priors for $\Omega_{m0}$ from Table \ref{table1x3}. Contour plots in the $\omega_{0}$-$\Omega_{m0}$ plane (right panel) with delta priors for $\omega_{1}$ from Table \ref{table1x3}.}
\label{fig1x3}
\end{figure}

\begin{figure}[!h]
\centering
\includegraphics[width=0.5\textwidth]{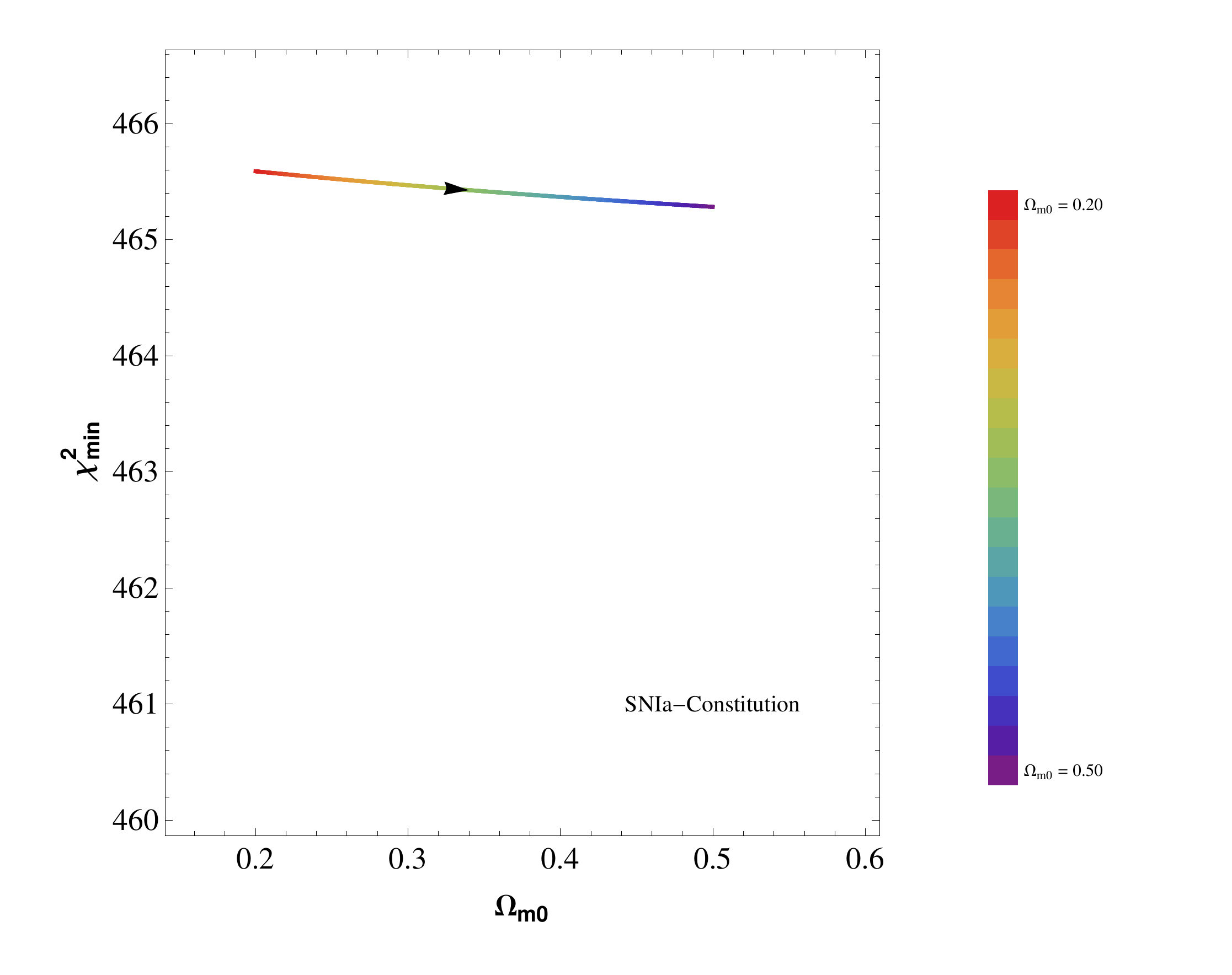}	
\includegraphics[width=0.38\textwidth]{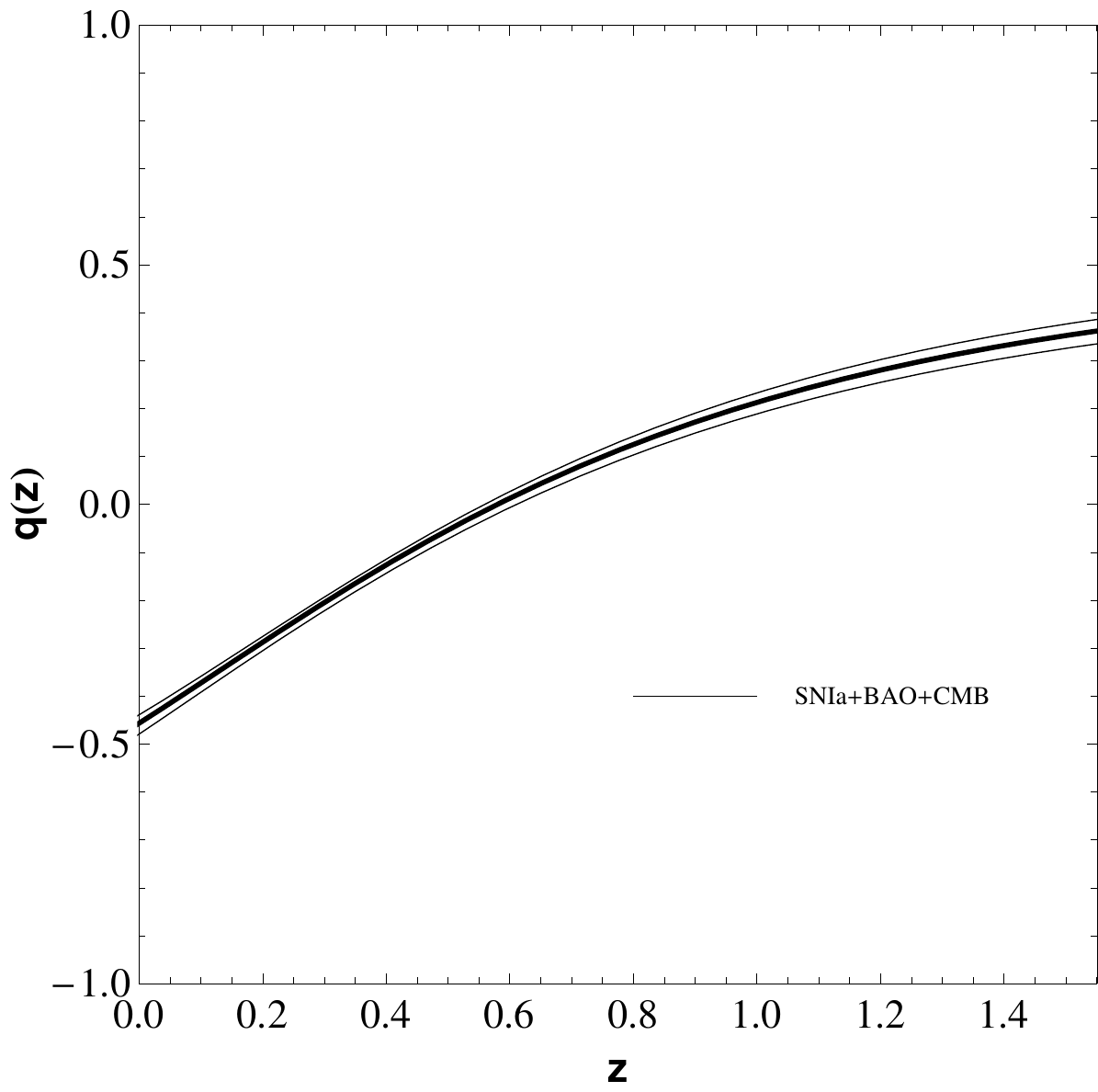}	
\caption{Case $\xi=3$. Left panel: dependence of $\chi^{2}_{min}$ on $\Omega_{m0}$.
 Notice that also here $\chi^{2}_{min}$ remains practically independent of $\Omega_{m0}$. Right panel: dependence of the deceleration parameter on the redshift (with $1\sigma$ contour lines) for  the   combination SNIa + BAO + CMB.}
\label{fig2x3}
\end{figure}

This model is characterized by Figs.~\ref{fig1x3} and \ref{fig2x3} as well as by tables~\ref{table1x3} and \ref{table2x3}.
Again, the minimum value of $\chi^{2}$ for the SNIa samples varies only slowly with $\Omega_{m0}$ with  a slight tendency toward  $\Omega_{m0}=1$ (left panel of Fig.~\ref{fig2x3}). Therefore, the
first rows in tables~\ref{table1x3} and \ref{table2x3} remain empty since no reliable entries are available.
For the same reason there do not appear confidence contours for the SNIa samples only in Fig.~\ref{fig1x3}. In the second rows of tables~\ref{table1x3} and \ref{table2x3} we include the parameter values for the priors $\Omega_{m0}=0.289$ and $\Omega_{m0}=0.269$, respectively.
The preferred $\omega_{1}$-values  are closer to the non-interacting case than to the case $\xi=1$ (left panel of Fig.~\ref{fig1x3}). Again, an insert magnifies the relevant region. Also the elongation in the right panel of Fig.~\ref{fig1x3} is reduced compared to its $\xi=1$ counterpart and, as in the non-interacting case, $\omega_{0} > -1$ is preferred.

At this place a comment on the redshift dependence of the deceleration parameter in the right panels of Figs.~\ref{figni2}, \ref{fig2x1} and \ref{fig2x3} is in order. According to the joint analysis, $q(z)$ decreases with decreasing $z$ in all the cases.  This contrasts with the
results found in \cite{sastaro}, according to which there exists a minimum in $q(z)$ which would imply
a slow down of the accelerated expansion and, possibly, a transition back to decelerated expansion.
However, as Fig.~\ref{figni2} also shows, considering only the SN data, the non-interacting model shows the same slowing-down behavior as that found in \cite{sastaro}.
From the Union2 data alone, a tendency that the cosmic acceleration will slow down was also reported
in \cite{caituo}.
For a combination of SNIa with the BAO data $D_{v}(z=0.35)/D_{v}(z=0.20)$ this result was confirmed in \cite{sastaro}.
For an analysis for $D_{v}(z=0.35)$, however, we were unable to reproduce this minimum. Also, if  CMB data are additionally included, the mentioned tendency for $q(z)$ to grow with decreasing redshift close to $z=0$ disappears.
A detailed analysis of this type of tensions has recently been performed in \cite{LiWuYu}.
A discussion on different kinds of tensions, including those between SNIa and CMB data, can be found, e.g., in \cite{haowei1,haowei2}.

\begin{table}[ht]
\centering  
\begin{tabular}{l  l  l  l  l  l  l} 
\hline\hline                        
\makebox [4 cm ][l]{Observation} &\makebox [2 cm ][l]{$\chi^2_{min}$} & \makebox [2 cm ][l]
{$\Omega_{m0}$} & \makebox [2 cm ][l]{$\omega_0$} & \makebox [2 cm ][l]{$\omega_1$} & \makebox [2 cm ][l]{$q_0$} \\ [0.5ex] 
\hline\hline                   
SNIa &   **  &  ** &   ** &  ** & **\\
(Prior $\Omega_{m0}=0.289$)&465.482  &  0.289 &-0.971  & -0.187 &-0.345 \\
\hline
SNIa+BAO & 465.495 &0.277 & -0.958 & -0.126 &-0.414\\
SNIa+CMB & 465.497 & 0.277  & -0.961 & -0.093&-0.451\\
SNIa+BAO+CMB & 465.497 & 0.277 & -0.963& -0.088&-0.457 \\ [1ex]      
\hline\hline  
\end{tabular}
\caption{Best-fit values based on the Constitution set for the case $\xi=3$ with the Hubble rate (\ref{H3}).}
\label{table1x3}
\end{table}

 \begin{table}[ht]
\centering
\begin{tabular}{l  l  l  l  l  l  l} 
\hline\hline                        
\makebox [4 cm ][l]{Observation} &\makebox [2 cm ][l]{$\chi^2_{min}$} & \makebox [2 cm ][l]
{$\Omega_{m0}$} & \makebox [2 cm ][l]{$\omega_0$} & \makebox [2 cm ][l]{$\omega_1$} & \makebox [2 cm ][l]{$q_0$} \\ [0.5ex] 
\hline\hline                   
SNIa &   ** &  ** & ** & ** & **\\
(Prior $\Omega_{m0}=0.269$)&  540.996   &  0.269 &-1.017  & 0.105 &-0.717 \\
\hline
SNIa+BAO & 540.994 &0.272 & -1.020 & 0.083 &-0.694\\
SNIa+CMB & 540.994 & 0.272  & -1.022& 0.101 &-0.715 \\
SNIa+BAO+CMB & 540.997 & 0.272 & -1.018  & 0.092&-0.702 \\ [1ex]      
\hline\hline  
\end{tabular}
\caption{Best-fit values based on the Union2 set for the case $\xi=3$ with the Hubble rate (\ref{H3}).}
\label{table2x3}
\end{table}

\subsubsection{Comparing the models}

\begin{figure}[!h]
\centering
\includegraphics[width=1.0\textwidth]{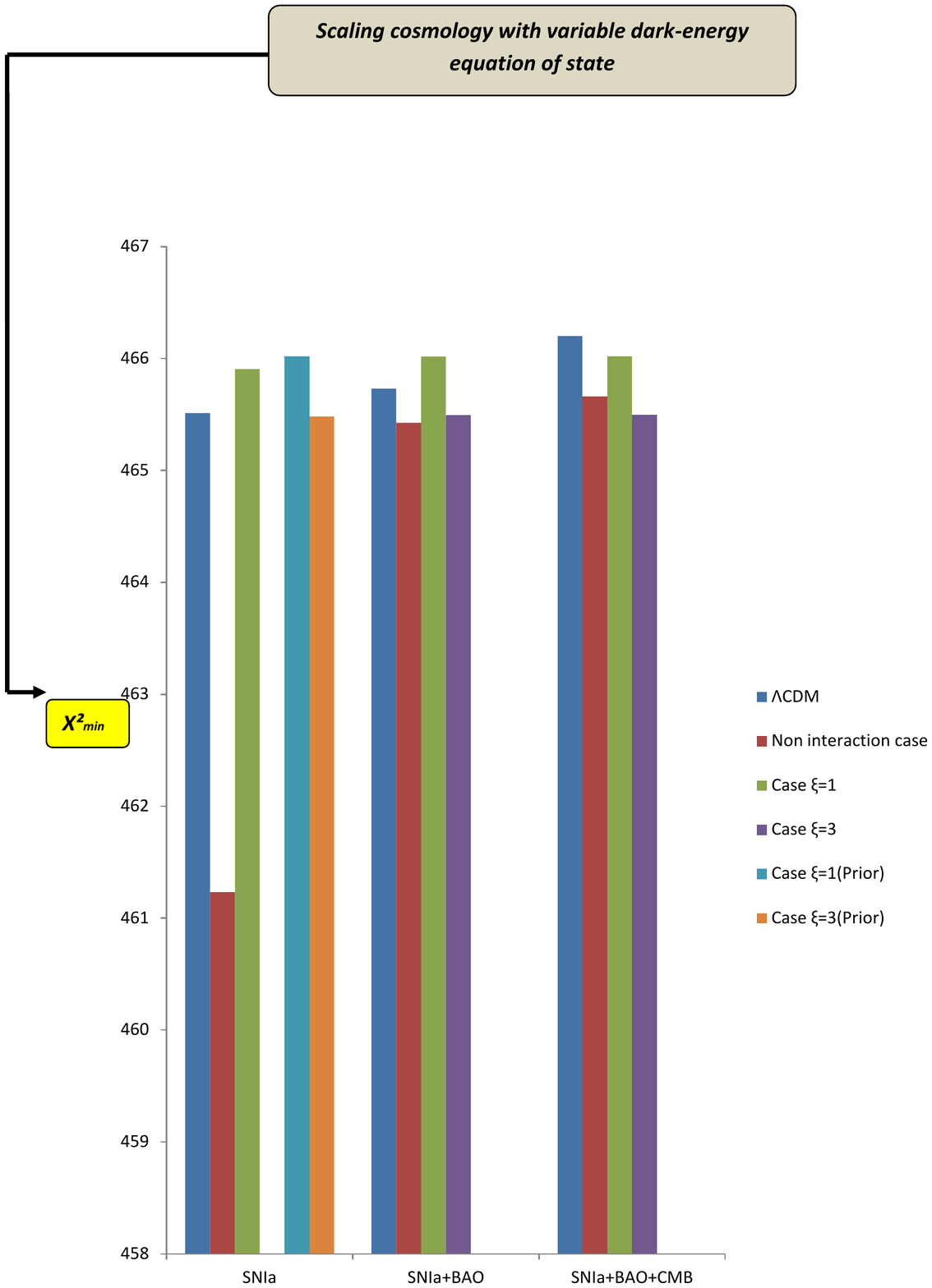}
\caption{ Graphical summary I of the analysis.}
\label{figsummary1}
\end{figure}

\begin{figure}[!h]
\centering
\includegraphics[width=0.8\textwidth]{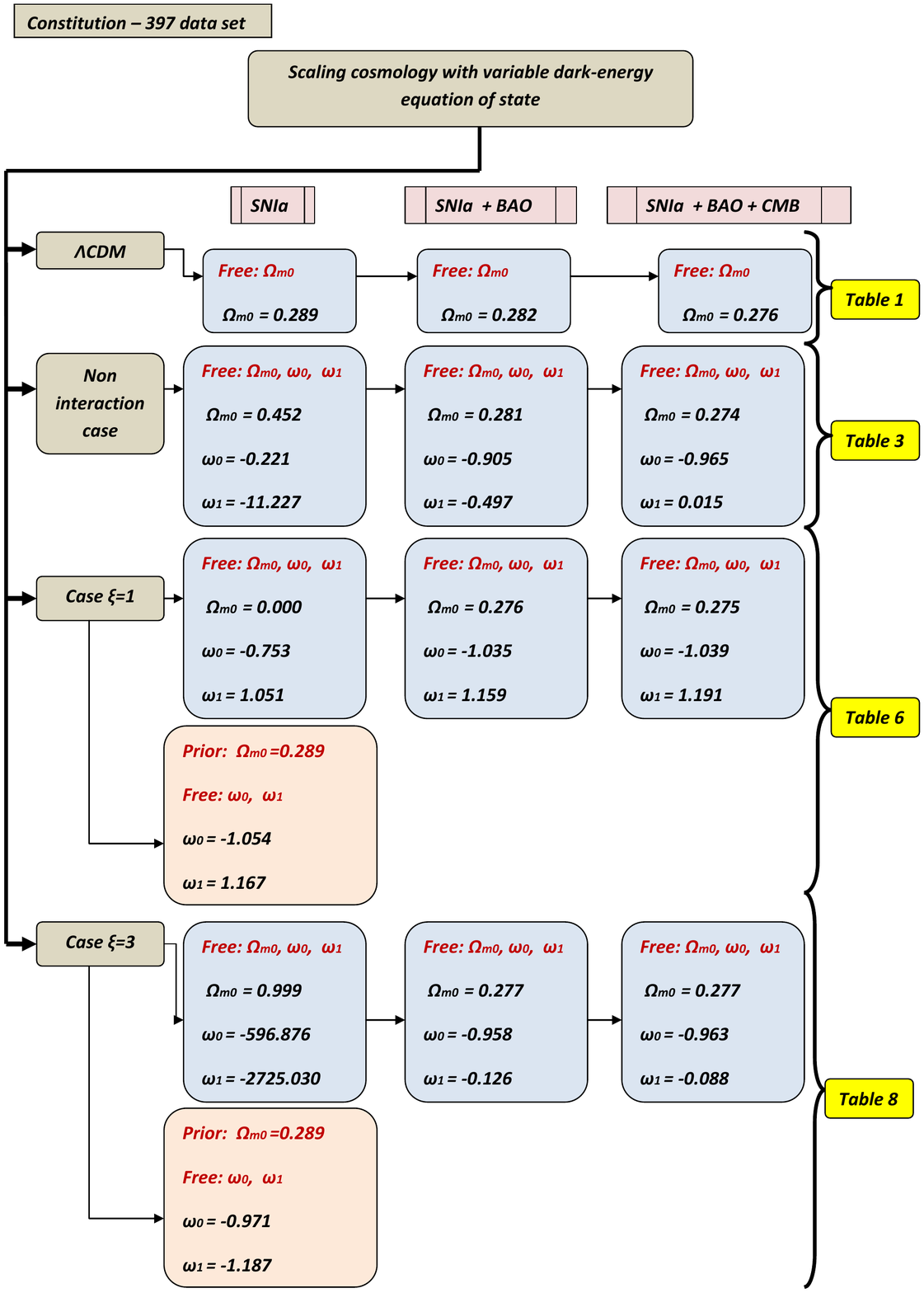}
\caption{Graphical summary II of the analysis.}
\label{figsummary2}
\end{figure}

The different models may be properly compared among themselves and also with the $\Lambda$CDM reference model, using appropriate statistical criteria. Two options are the already mentioned $AIC$ and $BIC$ criteria, which allow us to compare models with a different number of degrees of freedom.
The $AIC$ criterion uses the formula $AIC = \chi^2_{min} + 2k$ \cite{akaike}, where $k$ is the number of degrees of freedom; the
$BIC$ criterion \cite{schwarz} is based on the expression $BIC = \chi^2_{min} + 2k\ln N$, where $N$ is the number of observational points. The smaller the resulting numbers in both expressions, the higher the quality of the corresponding model.
It is convenient to classify a model with respect to the differences $\Delta$AIC and $\Delta$BIC between its $AIC$ and $BIC$ values, respectively, and the corresponding values for a reference model. This establishes a scale which allows for a ranking of different models according to the magnitude of their differences  $\Delta$AIC and $\Delta BIC$ \cite{liddle,mukh}. The smaller the difference to the lowest $AIC$ or $BIC$ values, here those of the $\Lambda$CDM model, the better the model. For differences less than $2$, there is strong support for the model under consideration.  If $\Delta AIC (\Delta BIC) < 6$ the model is still weakly supported. Models with $\Delta AIC (\Delta BIC) > 10$ should be considered as strongly disfavored.
In tables \ref{tablecritconst} and \ref{tablecritunion} we summarize the results for the investigated models and assess them according to the AIC and BIC criteria.
 Notice that all the $\chi^{2}_{min}$ values for the competing models are smaller than that of the $\Lambda$CDM model. But the mentioned criteria penalize the introduction of additional parameters and reverse the ranking.
By inspection, it follows, that, using the $AIC$ criterion, the non-interacting model and the $\xi = 3$ model are still weakly supported.
Applying, however, the $BIC$ criterion, all these models are ruled out.
This kind of contradiction in using
different evaluation criteria is well known in the literature, see, e.g., \cite{syz}.
Graphical summaries of our analysis is given in Figs.~\ref{figsummary1} and \ref{figsummary2}.

\begin{table}[ht]
\centering
\begin{tabular}{l  l  l  l  l  l} 
\hline\hline                        
\makebox [2 cm ][l]{Model}    &\makebox [2.5 cm ][l]{$\Lambda$CDM}      &\makebox [2.5 cm ][l]{No Int.}              & \makebox [2.5 cm ][l]{Int. $\xi=1$}       & \makebox [2.5 cm ][l]{Int. $\xi=3$}       \\ [0.5ex] 
\hline                   
Best fit &$\Omega_{m0} = 0.276$&$\Omega_{m0} =  0.274$  & $\Omega_{m0} = 0.275$ & $\Omega_{m0} = 0.277$  \\
         &                  &$\omega_0 = -0.965$ & $\omega_0 = -1.039$& $\omega_0 = -0.963$ \\
         &                  &$\omega_1 =  0.015$  & $\omega_1 = 1.191$ & $\omega_1 = -0.088$  \\
\hline
$q(z=0)$ &$q_0=-0.585$&$q_0=-0.550$  & $q_0=-0.630$ & $q_0=-0.457$\\
\hline
$\chi^2_{min}$ & 466.202 &465.606& 466.020 & 465.497 \\
$k$ & 1 & 3  & 3 & 3\\
$\Delta$BIC & 0 & 11.372 & 11.786& 11.263 \\
$\Delta$AIC & 0 & 3.404 & 3.818& 3.295 \\ [1ex]      
\hline\hline  
\end{tabular}
\caption{Summary of the analysis for the Constitution data set.}
\label{tablecritconst}
\end{table}

\begin{table}[ht]
\centering
\begin{tabular}{l  l  l  l  l  l} 
\hline\hline                        
\makebox [2 cm ][l]{Model}    &\makebox [2.5 cm ][l]{$\Lambda$CDM}      &\makebox [2.5 cm ][l]{No Int.}              & \makebox [2.5 cm ][l]{Int. $\xi=1$}       & \makebox [2.5 cm ][l]{Int. $\xi=3$}       \\ [0.5ex] 
\hline                   
Best fit &$\Omega_{m0} = 0.268$&$\Omega_{m0} =  0.269$  & $\Omega_{m0} = 0.270$ & $\Omega_{m0} =0.272 $  \\
         &                  &$\omega_0 = -1.030$ & $\omega_0 = -1.081$& $\omega_0 = -1.018$ \\
         &                  &$\omega_1 =  0.187$  & $\omega_1 = 1.269$ & $\omega_1 =0.092 $  \\
\hline
$q(z=0)$ &$q_0=-0.598$&$q_0=-0.630$  & $q_0=-0.683$ & $q_0=-0.702$\\
\hline
$\chi^2_{min}$ & 541.156 &541.070& 541.300 &540.997  \\
$k$ & 1 & 3  & 3 & 3\\
$\Delta$BIC & 0 & 12.559 & 12.789& 12.486 \\
$\Delta$AIC & 0 & 3.914 & 4.144 &  3.841\\ [1ex]      
\hline\hline  
\end{tabular}
\caption{Summary of the analysis for the Union2 data set.}
\label{tablecritunion}
\end{table}

\begin{figure}[!h]
\centering
\includegraphics[scale=1.3]{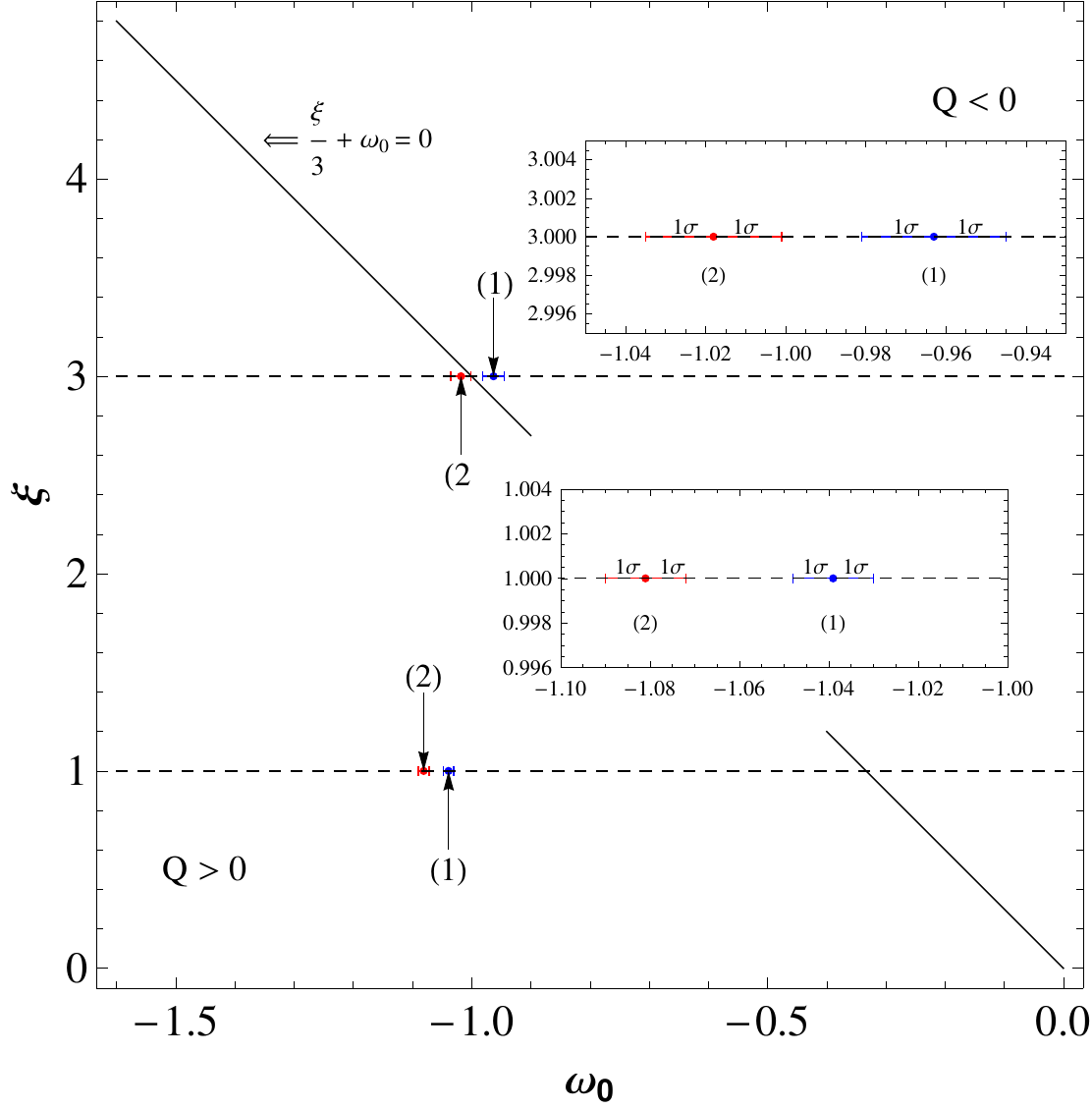}		
\caption{Direction of the energy transfer at the present epoch. The straight line divides the regions $Q>0$ and $Q<0$. The points denoted by $(1)$ correspond to the set
Constitution+BAO+CMB, the points denoted by $(2)$ correspond to Union2+BAO+CMB. The inserts magnify the $1\sigma$ regions.}
\label{figinter}
\end{figure}

Finally, we consider the direction of the energy transfer between the dark components.
 According to (\ref{Q}), the sign of $Q$ for the current Universe  depends on the sign of $\xi/3+\omega_0$. As to be seen in Fig.~\ref{figinter}, our interacting model with $\xi=1$ corresponds to $Q>0$ for both combinations of data sets. The case $\xi=3$ lies, as expected, very close to the non-interacting model. The Constitution data set yields $Q>0$ while the Union2 data set slightly favors an energy transfer from dark matter to dark energy, i.e, $Q<0$.

\section{Summary}
\label{discussion}

Cosmological models in which an interactions between dark matter and dark energy is admitted, give rise to a richer cosmological dynamics than non-interacting models, albeit at the expense of an additional parameter. The introduction of interacting models is largely motivated by the possibility to address the coincidence problem.
In this paper we have investigated a class of interactions that result in a power-law behavior
$\rho_{m}/\rho_{x}\propto a^{-\xi}$ of the ratio of the energy densities of dark matter and dark energy. Generalizing previous work, we admitted a time-varying EoS parameter of the dark-energy component within the CPL parametrization. We found analytic solutions for the cases $\xi = 1$ and $\xi = 3$.
The former is of interest with respect to an alleviation of the coincidence problem, the latter primarily to test a potential time variation of the EoS parameter.
With the help of a Bayesian statistical analysis we tested the resulting dynamics against the SNIa data
of the Constitution and Union2 samples. We included also information from BAO and CMB shift data and compared the interacting models among themselves and with a non-interacting model.
The $\chi^{2}_{min}$ values for all the competing models turned out to be smaller than the  $\chi^{2}_{min}$ value for the $\Lambda$CDM model. But according to both the $AIC$ and the $BIC$ criteria which penalize the introduction of additional parameters, the $\Lambda$CDM model remains the preferred choice.

Our study was restricted to the homogeneous and isotropic background dynamics. A more complete analysis requires to investigate the implications for structure formation as well. We believe, that the analytic solutions for the Hubble rates, found in this paper, will be helpful to calculate the matter power spectrum and the impact on the integrated Sachs-Wolfe effect in future work.

\acknowledgments
DRC is supported by CAPES. HV is supported by the CNPq (Programa Ci\^{e}ncia sem Fronteiras). WZ thanks CNPq for financial support.

\end{document}